\documentclass[preprint,amsmath,amssymb,phf,aip]{revtex4-1}
\usepackage{graphicx}
\usepackage{dcolumn}
\usepackage{bm}
\usepackage{float}
\begin{document}
\title{Dense granular flow of mixtures of spheres and dumbbells down a rough inclined plane: Segregation and rheology}
\author{Sandip Mandal}
\author{D. V. Khakhar}\email{khakhar@iitb.ac.in}
\affiliation{Department of Chemical Engineering, Indian Institute of Technology Bombay, Powai, Mumbai 400076, India}

\date{\today}

\begin{abstract}
We study the flow of equal-volume binary granular mixtures of spheres and dumbbells with different aspect ratios down a rough inclined plane, using the discrete element method. We consider two types of mixtures --  in the first type the particles of the two species have equal volume but different aspect ratios (EV) and in the second type they have variable volumes and aspect ratios (VV). We also use mixtures of spheres of two different sizes (SS) with the same volume ratios as in the mixtures of the second type, as the base case. Based on the study of Guillard, Forterre and Pouliquen [\textit{J. Fluid Mech.} \textbf{807}, R1--R11 (2016)], the inclination angle of the base for each mixture is adjusted and maintained at a high value to yield the same pressure and shear stress gradients for all mixtures and a high effective friction ($\mu$) for each. This ensures that the segregation force and resulting extent of segregation depend only the size and shape of the particles. The species with larger effective size, computed in terms of the geometric mean diameter, floats up in all cases and the dynamics of the segregation process for all the mixtures are reported. The concentration profiles of the species at steady state agree well with the predictions of a continuum theory. The $\mu-I$ and $\phi-I$ scaling relations, where $I$ is the inertial number and $\phi$ is the solid volume fraction, extended to the case of mixtures, are shown to describe the rheology for all the cases.     
 \end{abstract}

\maketitle

\section{Introduction}
The granular mixtures handled in industries, such as coke-limestone-iron ore mixtures in steel industries, powders in pharmaceutical industries, limestone-clay mixtures in cement industries, sand-gravel-cement mixtures in construction industries, etc., are composed of spherical and nonspherical particles. Such mixtures often exhibit demixing or segregation \cite{williams1976segregation,fan1990recent,ottino2000mixing,kudrolli2004size} during pouring, tumbling, shaking, and conveying, due to the species differing in size \cite{savage1988particle,khakhar1999mixing,thomas2000reverse,kudrolli2004size,jain2005regimes,gray2006particle,tripathi2011rheology,thornton2012modeling,fan2014modelling,gray2015particle,guillard2016scaling,gray2018}, density \cite{khakhar1997radial,khakhar1999mixing,jain2005regimes,sarkar2008experimental,tripathi2013density,Tunuguntla2014,Gray2015}, shape \cite{pouliquen1997fingering,grasselli1998experimental,makse1998spontaneous,pouliquen1999segregation, abreu2003influence,caulkin2010geometric,roskilly2010investigating,woodhouse2012segregation,yuan2013segregation,windows2015competition,baker2016segregation,windows2016modifying,pereira2017segregation,zhao2018simulation}, and mechanical properties, such as friction coefficient \cite{lai1997friction,hajra2004sensitivity,ulrich2007influence,dvziugys2009role,gillemot2017shear} and restitution coefficient \cite{brito2008segregation,windows2014effects,windows2014inelasticity}. The segregation is unwanted and needs in-depth understanding since it affects product quality \cite{gray2015particle}. 

When a granular mixture, composed of different size species, flows down an inclined plane, the larger species rises to the top, and the smaller species percolates to the bottom of the flowing layer as a result of \textit{kinetic sieving} and \textit{squeeze expulsion} \cite{savage1988particle,khakhar1999mixing,thomas2000reverse,gray2006particle,tripathi2011rheology,wiederseiner2011experimental,staron2014segregation,guillard2016scaling,Gray2015}.  When there is a density difference between the species, the heavier species sinks to the bottom displacing the lighter species to the top \cite{khakhar1999mixing,tripathi2011rheology,tripathi2013density,Gray2015}. In rotating cylinders, the smaller or denser species forms a core near the axis (center), and the larger or lighter species concentrates near the periphery \cite{nityanand1986analysis, khakhar1997radial, khakhar2003segregation,jain2005regimes,hill2008kinematics, sarkar2008experimental,hajra2005radial,hajra2011radial,khakhar2011rheology,hill2014segregation,schlick2015granular}. When a granular mixture of large nonspherical (or spherical) and small spherical beads is poured into a quasi-two-dimensional bin, it either stratifies into alternating layers of large and small beads spanning down the entire length of the flowing layer or purely segregates with the smaller beads near the top end (near the pouring point) and the larger ones near the bottom end of the heap \cite{williams1968,gray1997pattern,grasselli1998experimental,makse1998spontaneous,baxter1998stratification,makse1998dynamics,fan2012stratification,schlick2015modeling,fan2017segregation} depending on the size ratio and flow rate. When a bed of granular mixture is shaken, the larger species rises to the top and is termed as `Brazil nut effect' \cite{rosato1987brazil,knight1993vibration}. However, the reverse scenario is also observed depending on the density ratio of larger to smaller species \cite{shinbrot1998reverse,hong2001reverse,breu2003reversing}. 

The segregation induced by the difference in the shape of the species is less explored, and its mechanism is not yet fully understood. Mixtures comprising large nonspherical and small spherical beads stratify during heap formation due to the difference in the angle of repose \cite{gray1997pattern,grasselli1998experimental,makse1998spontaneous} between the species, as mentioned earlier. Similarly, mixtures comprising large angular and small spherical beads exhibit finger formation while flowing on a rough inclined plane \cite{pouliquen1997fingering,pouliquen1999segregation,woodhouse2012segregation,baker2016segregation}. The fingering instability is driven by the difference in the friction coefficients between the species. The tendency of minimizing the global potential energy (or porosity) of a system was shown to drive the shape induced segregation \cite{abreu2003influence}. The monodisperse spherocylinders of moderate aspect ratios (2.5$>$aspect ratio$>$1.125) created denser packing than the monodisperse spheres of the same volumes. Hence, when a mixture of spheres and spherocylinders was vertically shaken, the spherocylinders settled beneath the spheres and thereby forced the system to reach the minimum global potential energy state \cite{abreu2003influence}. However, if the species had different volumes, the \textit{percolation} effect dominated, and the species with higher volume always rose to the top irrespective of their shape \cite{abreu2003influence}. The difference in the effective size of the species, calculated based on the radii of gyration of the particles, was also shown to drive the segregation process \cite{roskilly2010investigating,windows2015competition}. The species with larger effective size always rose to the top of a vibrating bed irrespective of their shape \cite{roskilly2010investigating,windows2015competition}, provided the two species were of the same volume. Recently, \citet{pereira2017segregation} studied the segregation of binary granular mixtures of spheres and cubes of the same volume in a rotating tumbler. The cubes concentrated near the axis (center), and the spheres accumulated near the periphery. The segregation was driven by the difference in the flowability of the species in this case. 

The rheology of granular mixtures of spheres of different size/density or both has been studied to a greater intensity in chutes, rotating cylinders, heaps, and Couette shear cells \cite{hill2008kinematics,sarkar2008experimental,rognon2007dense,yohannes2010rheology,tripathi2011rheology,tripathi2013density,fan2013kinematics}. Both species reach a thermal equilibrium showing identical mean velocity, shear rate, and stress profiles \cite{hill2008kinematics,sarkar2008experimental,tripathi2011rheology,tripathi2013density,fan2013kinematics}.  The $\mu-I$ and $\phi-I$ scalings, where $\mu$ is the effective friction coefficient, $\phi$ is the volume fraction, and $I$ is the inertial number, with mixture rules are applicable for describing the rheology with the same model parameters as in the monodisperse case \cite{rognon2007dense,yohannes2010rheology,tripathi2011rheology}. Note that these constitutive scaling relations are well-posed in the dense flow regime for the moderate values of the inertial number \cite{barker2015well}. However, they are ill-posed in the quasi-static and collisional flow regimes for small and large values of the inertial number, respectively \cite{barker2015well}. A number of studies are directed toward making the equations well-posed across all the flow regimes \cite{kamrin2012nonlocal,bouzid2013nonlocal,barker2017well,heyman2017compressibility,goddard2017stability}.   

A study by \citet{guillard2016scaling} showed that the net segregation force ($F_{seg}$) experienced by a single large circular intruder of diameter $d_c$ in a flowing medium of small circular particles of diameter $d$ on an inclined plane may be expressed as  
\begin{equation}\label{eq:force}
F_{seg}=\pi\frac{d^2_c}{4}\Big(\textit{F}(\mu,d_c/d)\frac{\partial P}{\partial y}+\textit{G}(\mu,d_c/d)\frac{\partial |\tau_{xy}|}{\partial y}\Big),
\end{equation}  
where $\mu=|\tau_{xy}|/P$, $\partial P$/$\partial y$ is the gradient of pressure, $\partial |\tau_{xy}|$/$\partial y$ is the gradient of shear stress, and $\textit{F}(\mu,d_c/d)$ and $\textit{G}(\mu,d_c/d)$ are empirical functions satisfying $\textit{F}(\mu,1)=1/\phi(\mu)$ and $\textit{G}(\mu,1)=0$. Eq.~(\ref{eq:force}) indicates that the gradients of pressure and shear stress independently influence the upward and downward migration of the intruder, respectively. Further, \citet{guillard2016scaling} showed that the net segregation force was nearly independent of the effective friction ($\mu$) at its high values, but reduced sharply at low values. Eq.~(\ref{eq:force}) qualitatively described segregation of binary granular mixtures of spheres of different sizes in different geometries \cite{guillard2016scaling}. We assume in this study that the net segregation force given in Eq.~\ref{eq:force} is valid even for the case of binary mixtures of spheres and dumbbells with $F$, $G$ and the prefactor to be dependent on shape as well.    

In this study, we investigate the rheology and segregation of binary granular mixtures of spheres and dumbbells flowing down a rough inclined plane by using the soft sphere, discrete element method (DEM) simulations. Shape-induced segregation has practical applications but has not been explored in detail for steady flows as yet. Based on the findings of \citet{guillard2016scaling}, we carry out simulations in which the pressure and shear stress gradients are maintained the same for different mixtures, and a sufficiently high value of the shear rate is used at which the friction coefficient is high so that the segregation force depends only on the differences in the size and shape of the species. Our objectives are to study the effect of particle shape on segregation and to characterize the rheology of the system. The simulation methodology is explained in section~\ref{sec:simulation}. Simulation results are reported in section~\ref{sec:results and discussion}, followed by conclusions in section~\ref{sec:conclusions}.   

\section{Simulation methodology}\label{sec:simulation}
\begin{figure}
\centerline{\includegraphics[width=6.5in]{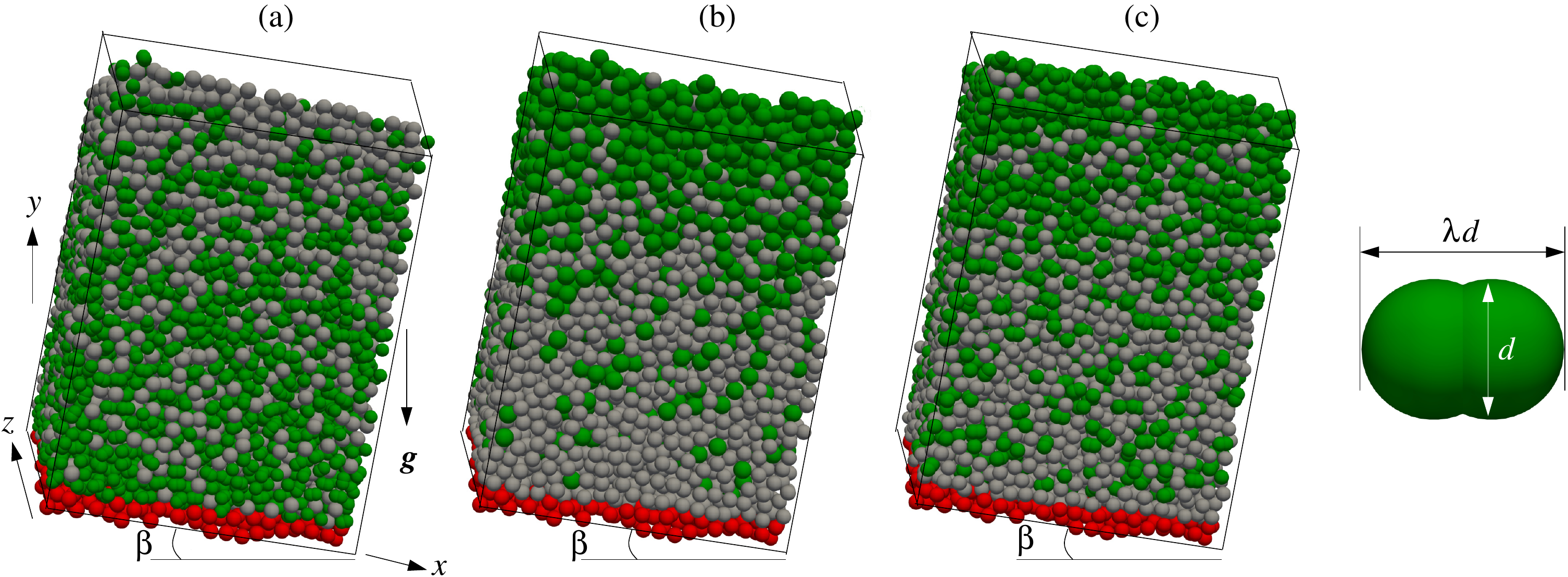}}
\caption{Snapshots of the system at steady state showing the flow of 50\% volume mixtures (a) \textit{EV2}, (b) \textit{SS4}, and (c) \textit{VV4}, as given in Table~\ref{tab:mixtures}. The dumbbells in green shade concentrate near the base in (a), however, float up near the free surface in (c). The larger spheres in green shade float up near the free surface in (b), as expected. The simulation box has a rough base (shown in red shade) and is inclined at $\beta=25.7 ^\circ$ in (a), 24.5$^\circ$ in (b), and 24.6$^\circ$ in (c). The flow occurs in the $x$ direction. Gravity ($\textbf{g}$) acts vertically downward. A dumbbell with its dimensions ($d$,$\lambda d$) is shown alongside.}
\label{fig:system}
\end{figure}

We simulate binary granular mixtures of frictional, inelastic spheres and dumbbells of different aspect ratios flowing down a bumpy inclined plane by using the soft sphere, discrete element method. The dumbbells are made by fusing two spheres of mass $m$ and diameter $d$ and removing the half the overlapping volume \cite{Mandal2D,Mandal3D} (Fig.~\ref{fig:system}, Table~\ref{tab:dumbbells}). The dumbbells closely resemble food grains \cite{mandal2017}. The aspect ratio ($\lambda$) of the dumbbells is increased two ways: (\textit{i}) by reducing the diameters of the component spheres and adjusting the center-to-center distance between the spheres keeping the particle volume fixed and (\textit{ii}) by increasing the center-to-center distance between the spheres keeping their diameters constant ($d=1$), which results in increasing particle volume (Table~\ref{tab:dumbbells}). The structural properties of the dumbbells are tabulated in Table~\ref{tab:dumbbells}. We define the effective size of the particles in terms of the radius of gyration $r_g=\sqrt{(I_p/m_p)}$, where $I_p$ is the moment of inertia, and $m_p$ is the mass of the particles and in terms of the geometric mean diameter $d_g=\lambda^{1/3}d$. Two types of mixtures are used: \textit{Equal Volume} (\textit{EV}) mixtures comprising species with the same volume but different aspect ratios and \textit{Varying Volume} (\textit{VV}) mixtures comprising species with different volumes and aspect ratios (Table~\ref{tab:mixtures}). Both species have the same density ($\rho_p$), and the particles are of identical size (monodisperse) in each case. We also use mixtures comprising two different sized spheres (\textit{Spheres-Spheres} (\textit{SS})) with the same volume ratios as in the \textit{VV} mixtures (Table~\ref{tab:mixtures}) for comparison. The species with either larger aspect ratio or larger size is referred to as species $B$ in each case. The simulation box, shown in Fig.~\ref{fig:system}, is 20 units long and wide in the flow ($x$) and vorticity ($z$) directions. Periodic boundary conditions are applied in both flow and vorticity directions to eliminate the end and side wall effects. A 1.2 units thick slice of a random close packed bed of dumbbells of aspect ratio $\lambda=1.0005$ is glued on a flat surface to generate the bumpy base. 

\begin{table}
\caption{\label{tab:dumbbells} Structural properties of the dumbbells.}
\begin{ruledtabular}
\begin{tabular}{ccccccc}
Case  & Component  & Particle  & Aspect & Radius of  & Geometric mean & Particle \\
            & diameter ($d$) & volume ($v_p$) & ratio ($\lambda$) & gyration ($r_g$) & diameter ($d_g$) & shape\\[1pt]
\colrule
 & 1.0 & 0.524 & 1.0 & 0.316 & 1.0 &\parbox[c]{1em}{\includegraphics[width=9.2mm]{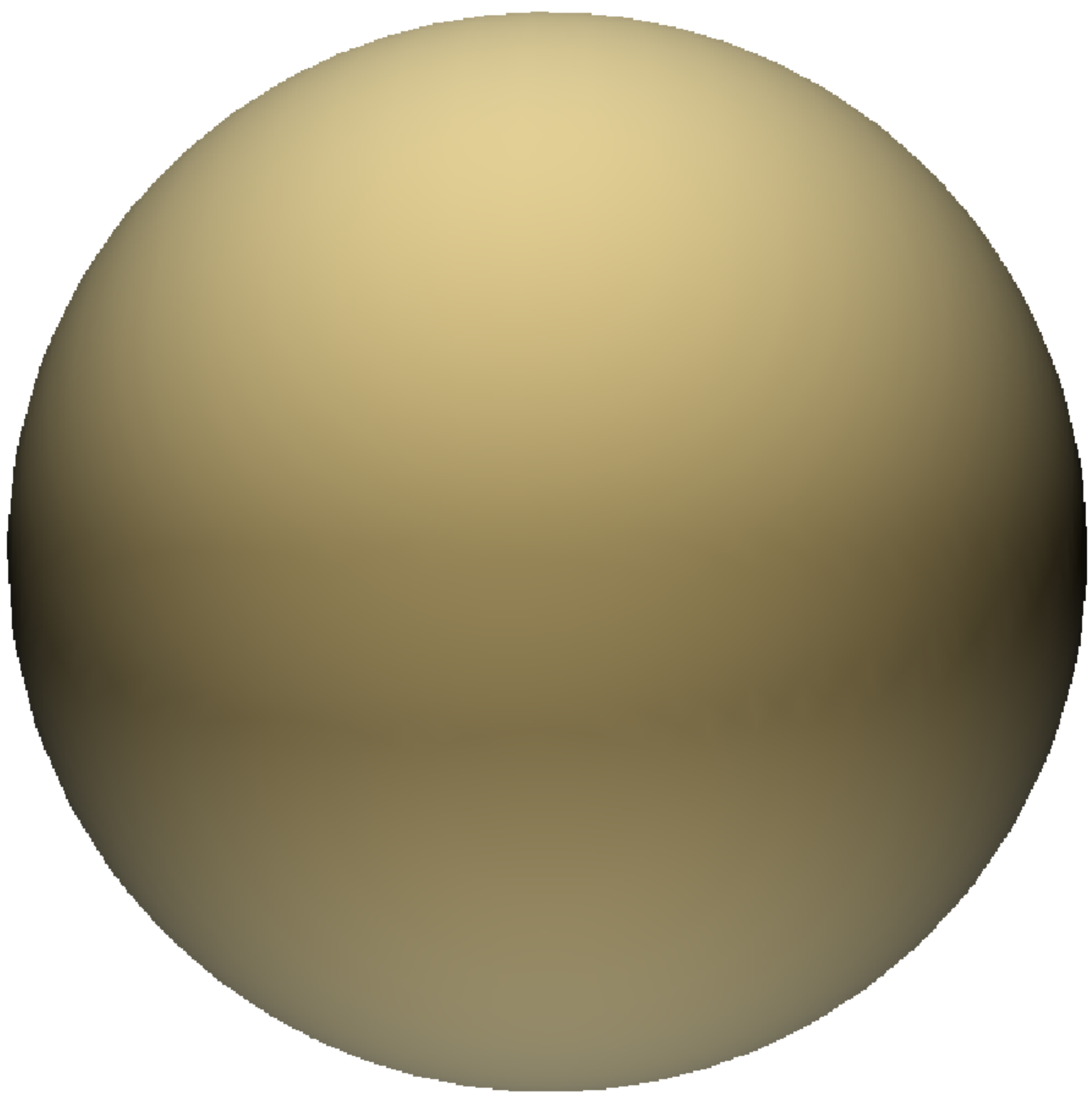}} \\
      varying & 0.92 & 0.526 & 1.20 & 0.284 & 0.978 &\parbox[c]{1em}{\includegraphics[width=10.0mm]{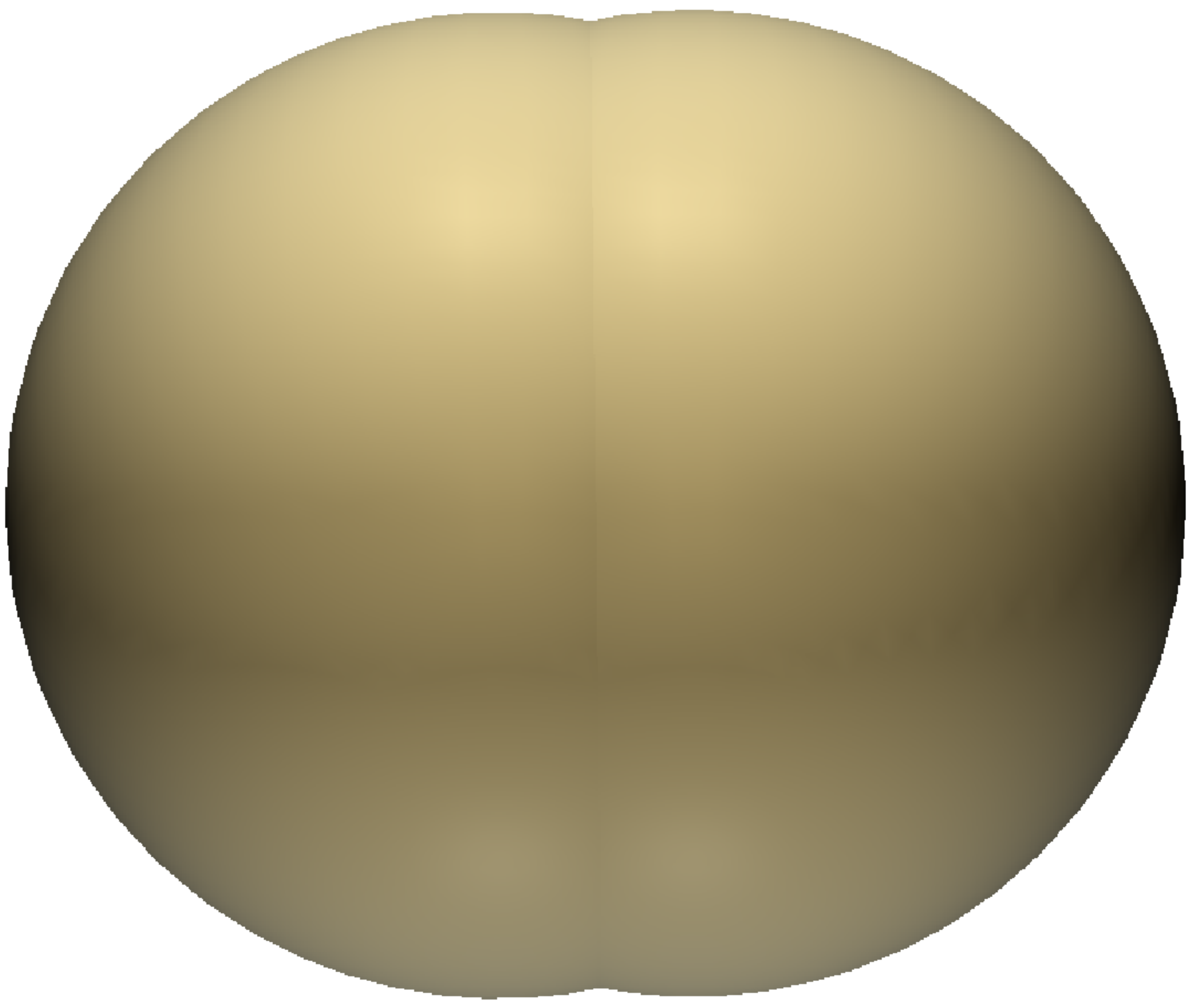}}\\
      component & 0.86 & 0.522 & 1.40 & 0.284 & 0.962 &\parbox[c]{1em} {\includegraphics[width=10.8mm]{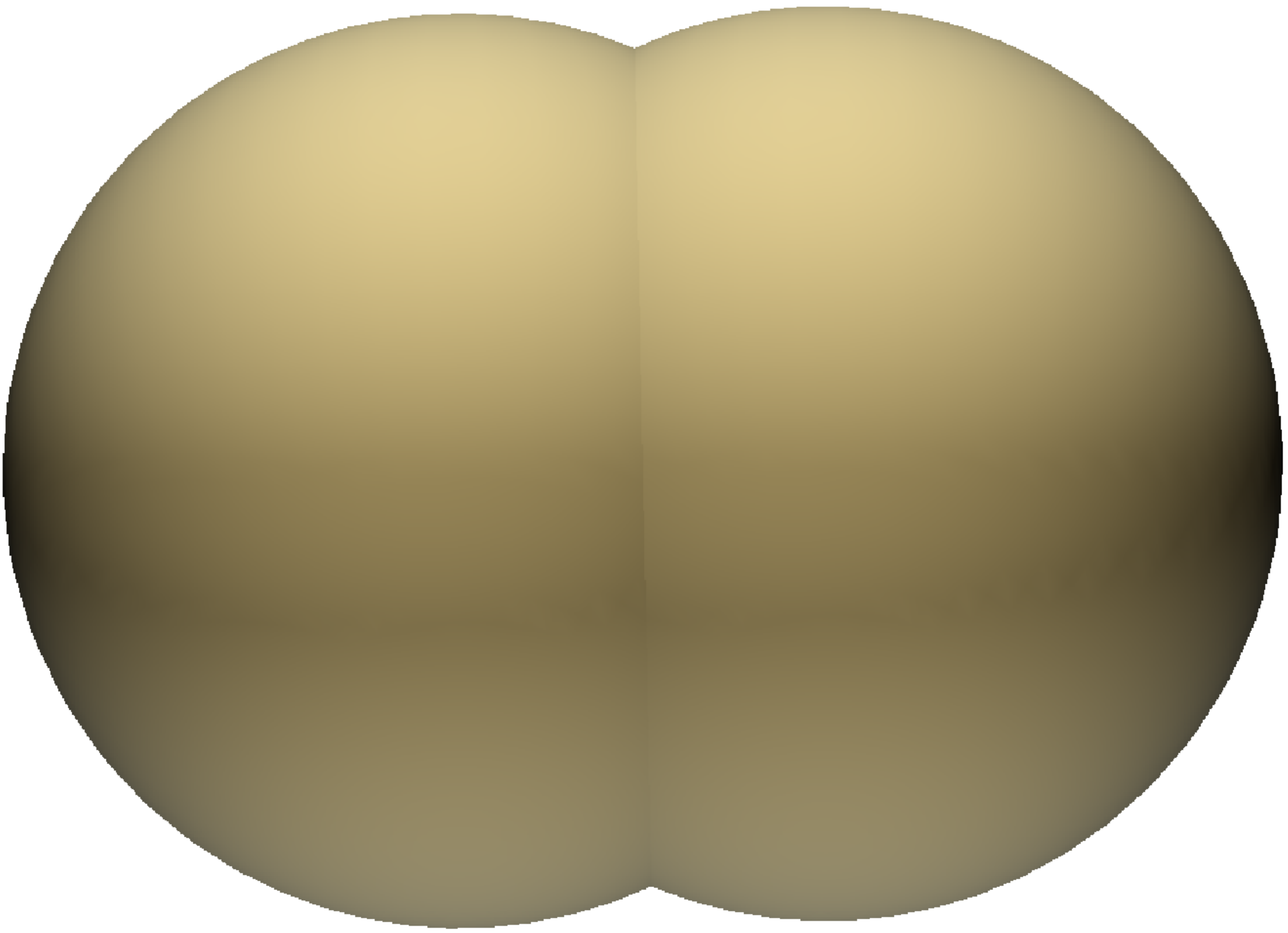}}\\
      diameter & 0.831 & 0.523 & 1.55 & 0.299 & 0.962 &\parbox[c]{1em} {\includegraphics[width=11.6mm]{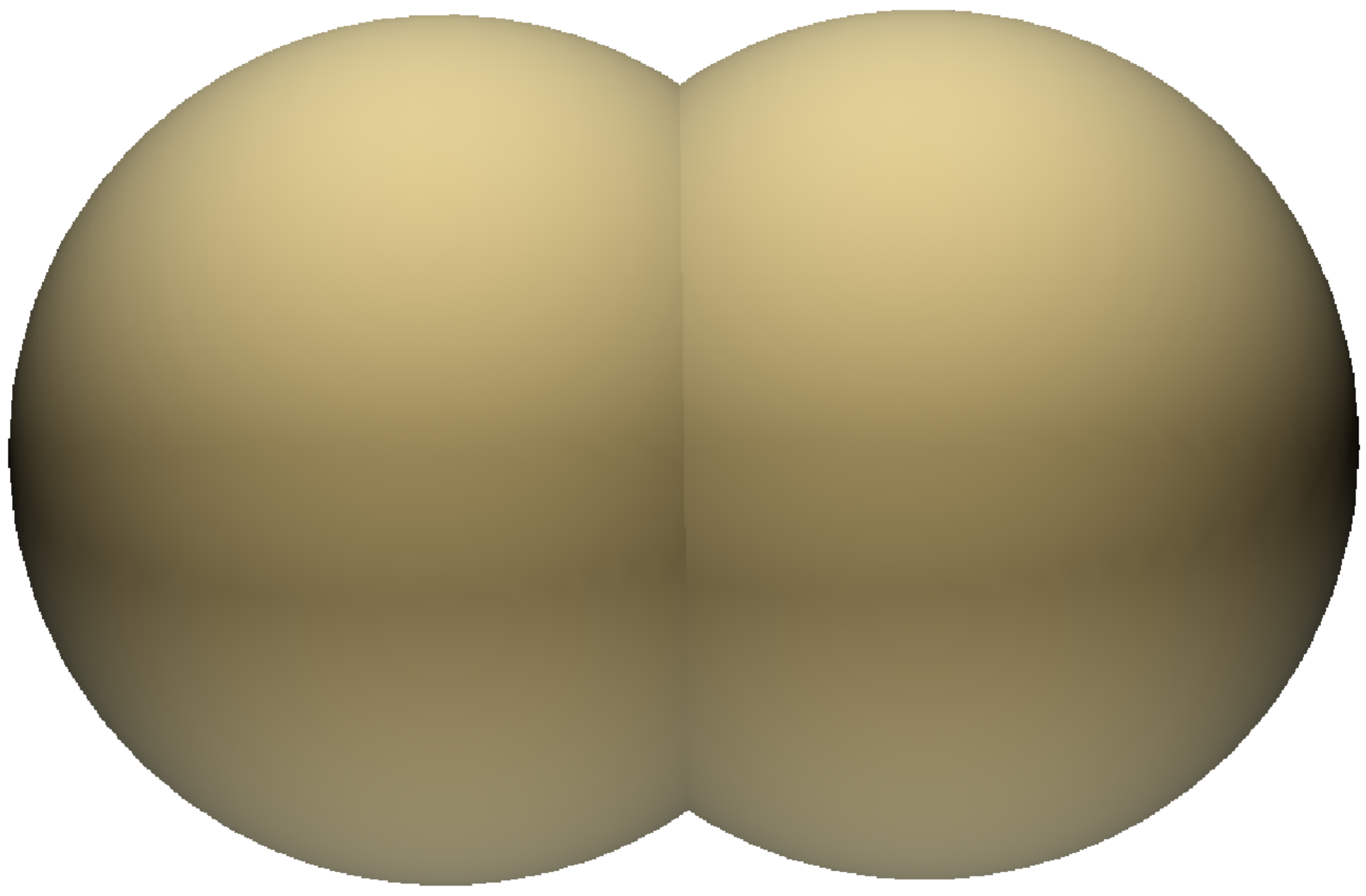}}\\
      & 0.81 & 0.523 & 1.70 & 0.319 & 0.967 &\parbox[c]{1em} {\includegraphics[width=12.2mm]{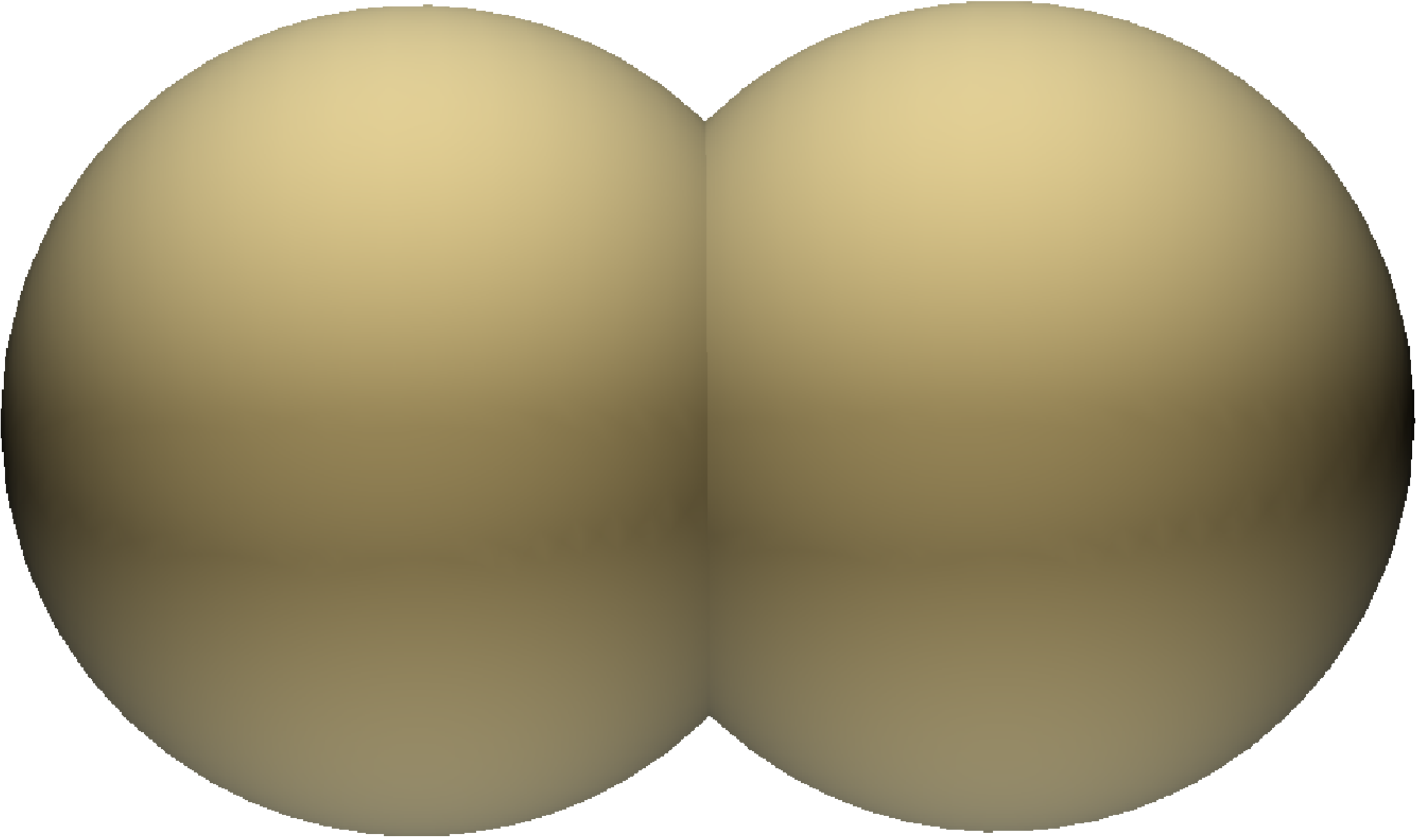}}\\
      \hline
      & 1.0 & 0.524 & 1.0 & 0.316 & 1.0 &\parbox[c]{1em}{\includegraphics[width=9.2mm]{sphere}} \\
       fixed & 1.0 & 0.602 & 1.1 & 0.308 & 1.032 &\parbox[c]{1em} {\includegraphics[width=10.2mm]{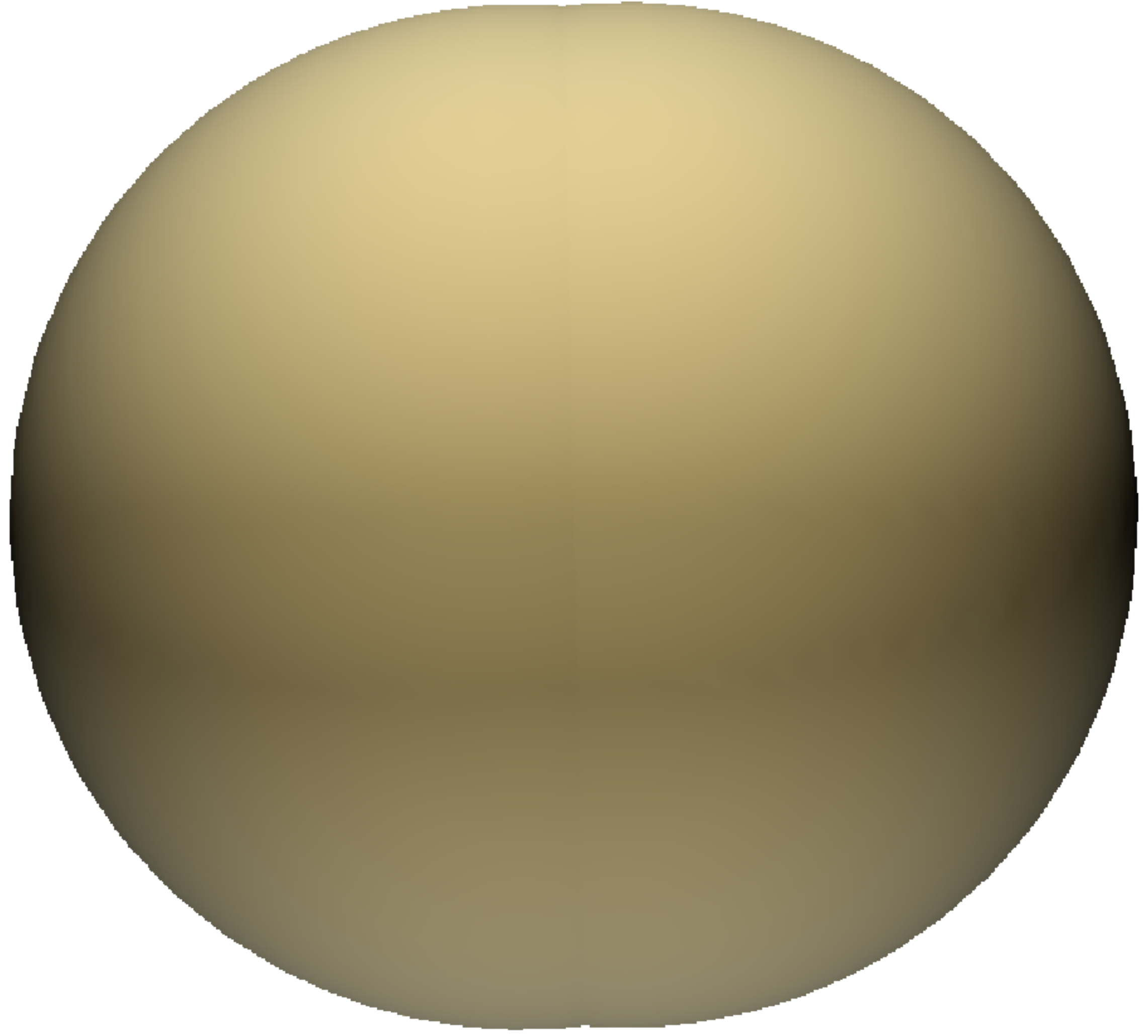}}\\
      component & 1.0 & 0.679 & 1.2 & 0.308 & 1.063 &\parbox[c]{1em} {\includegraphics[width=11.0mm]{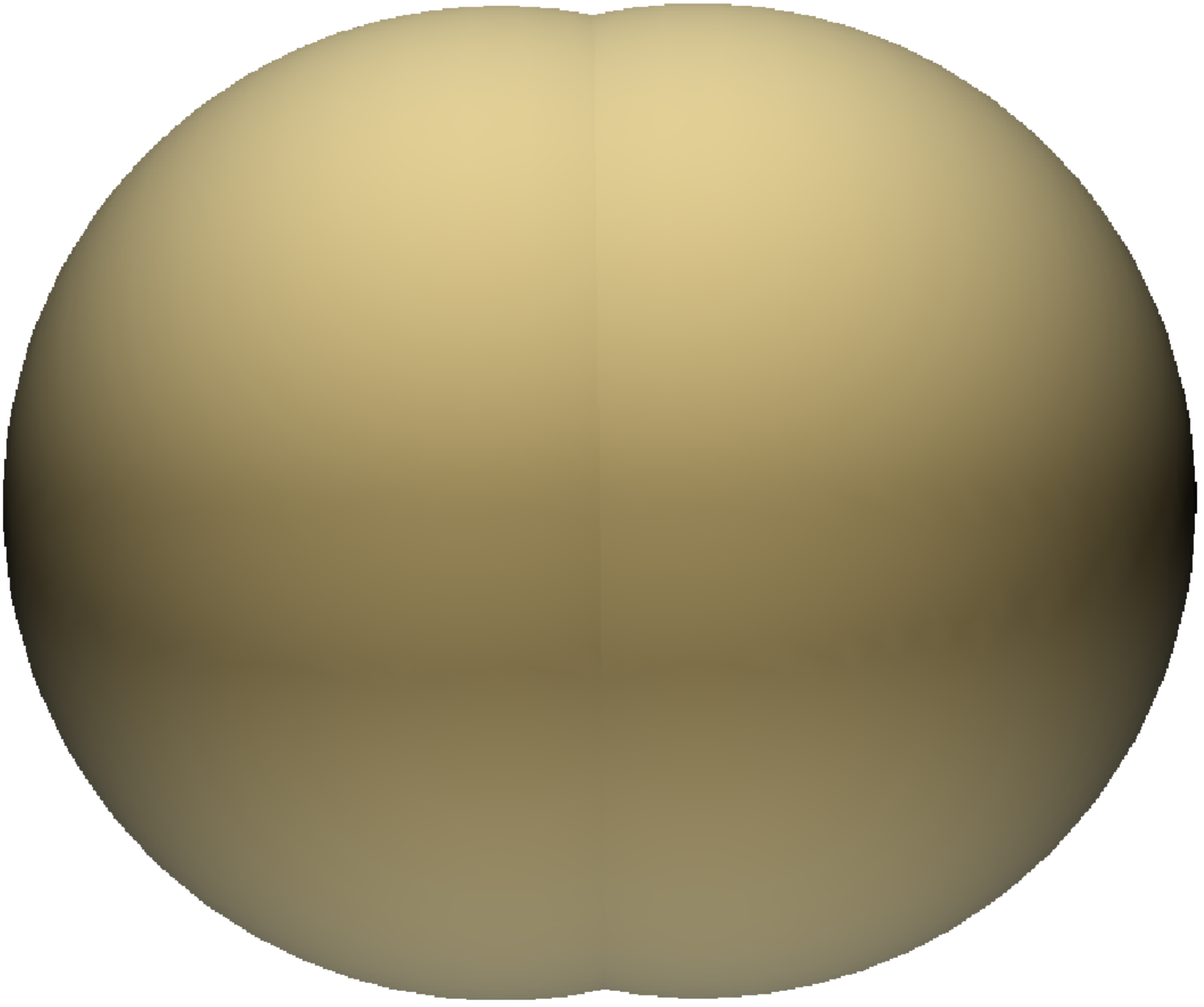}} \\
      diameter & 1.0 & 0.752 & 1.3 & 0.316 & 1.091 &\parbox[c]{1em} {\includegraphics[width=11.8mm]{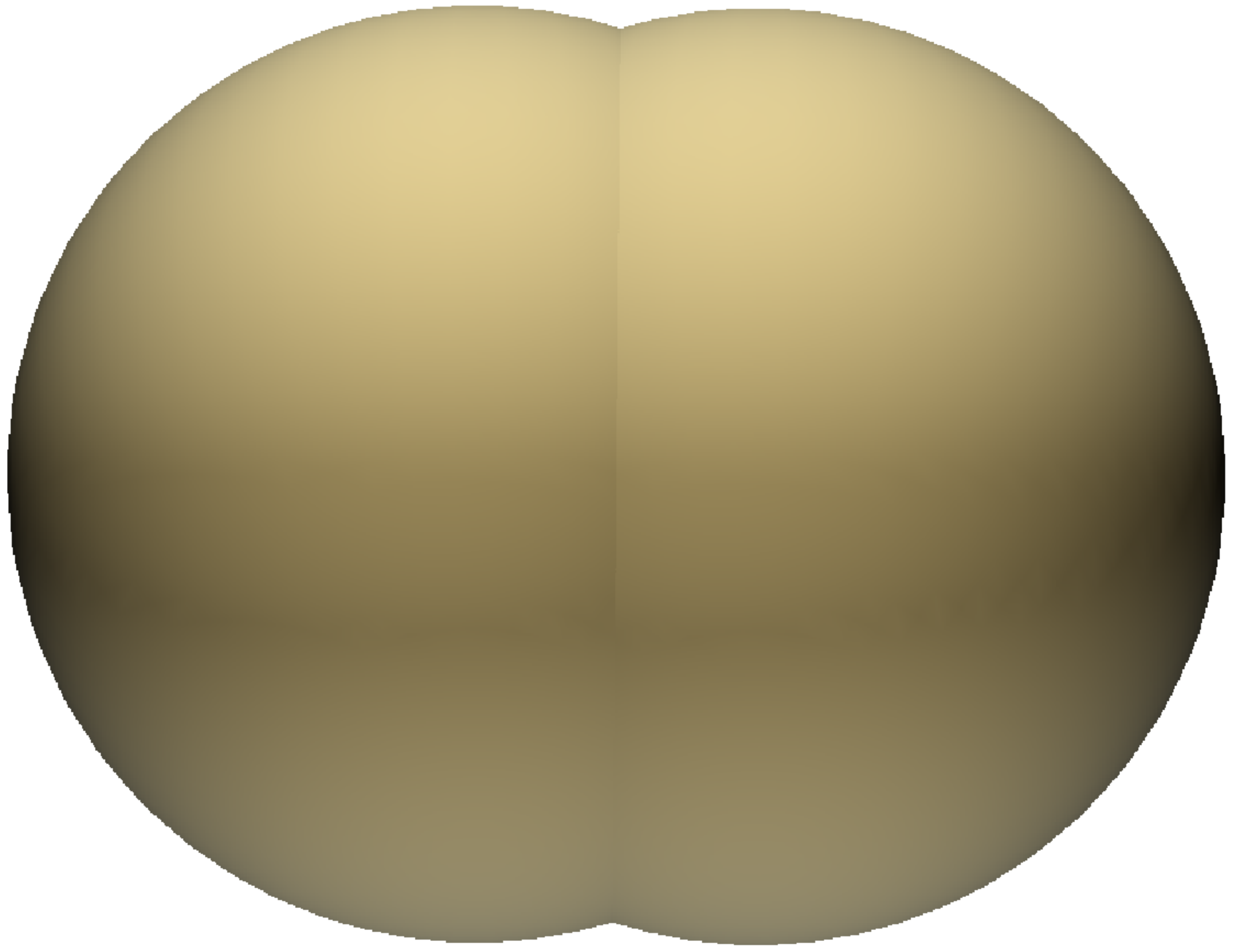}}\\
      & 1.0 & 0.821 & 1.4 & 0.33 & 1.119 &\parbox[c]{1em} {\includegraphics[width=12.5mm]{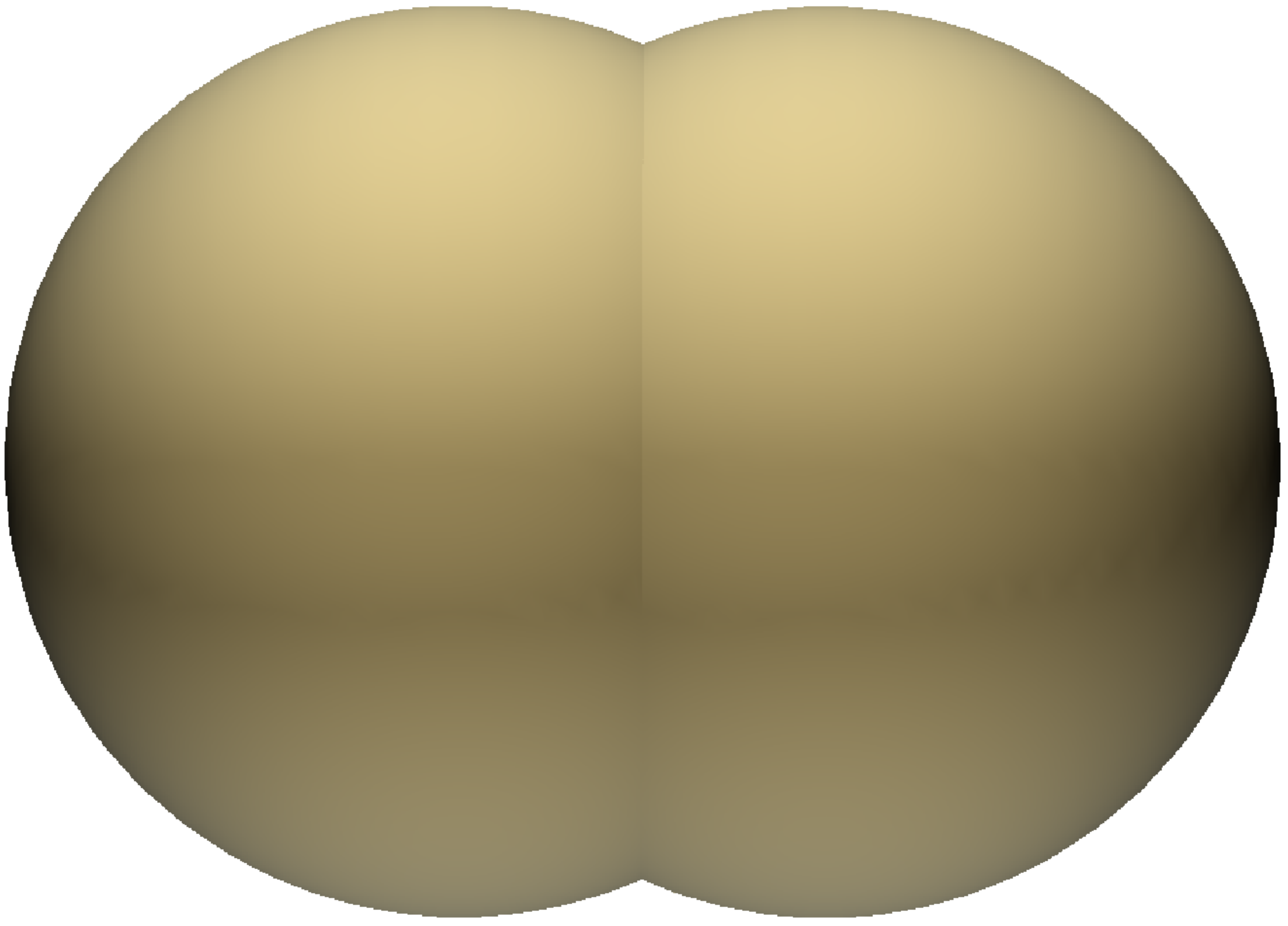}}\\
\end{tabular}
\end{ruledtabular}
\end{table}

The contact forces are calculated using a linear spring-dashpot model with a frictional slider, which is identical to the $L3$ model of \citet{silbert2001granular}. The contact forces, generated due to the deformations of the spheres, comprise elastic spring forces which try to restore the deformations, and viscous damping forces which drain out the energy of inelastic collisions. We sum up the forces and torques acting on the two spherical components to get the total force and torque acting on a dumbbell \cite{Mandal2D,Mandal3D}. The equations of motion are non-dimensionalized using $d$, $m$, $mg$, and $\sqrt{d/g}$ as length, mass, force, and time units, respectively and are integrated using the Verlet algorithm with a time step $dt$ $\approx$ 10$^{-4}$ to get the updated positions and velocities of the particles. The DEM code is validated by comparing the mean velocity and volume fraction profiles of monodisperse dumbbells of a very small aspect ratio ($\lambda=1.0005$) with those of the monodisperse spheres in the studies of \citet{silbert2001granular} and \citet{tripathi2011rheology}, using the same model parameters, as given below. In our simulations, the particles have coefficient of restitution $e=0.88$, coefficient of sliding friction $\mu_p=0.5$, dimensionless normal spring constant $k_n=2\times10^5$, and dimensionless tangential spring constant $k_t=2/7 \times k_n$, which correspond to those used in the $L3$ model of \citet{silbert2001granular}. The detailed simulation methodology is given elsewhere \cite{Mandal2D}.

\begin{table}[ht!]
\caption{\label{tab:mixtures} Composition of the mixtures.}
\begin{ruledtabular}
\begin{tabular}{cccc@{\hspace{0.6cm}}cccc}
Type & Mixture & Species $A$ & Species $B$ & $\lambda^A$  & $\lambda^B$ & $d^B_g/d^A_g$ & $r^B_g/r^A_g$ \\[1pt]
      \hline
       & \textit{EV1}  & \parbox[c]{1em}{\includegraphics[width=9.2mm]{sphere}} &\parbox[c]{1em}{\includegraphics[width=10.0mm]{dumbbell-12}} & 1.0 & 1.20 & 0.98 & 0.89\\
      \textit{Equal}& \textit{EV2} & \parbox[c]{1em}{\includegraphics[width=9.2mm]{sphere}} & \parbox[c]{1em} {\includegraphics[width=10.8mm]{dumbbell-14}} & 1.0 & 1.40 & 0.96 & 0.89\\
     \textit{Volume} (\textit{EV})  & \textit{EV3} & \parbox[c]{1em}{\includegraphics[width=9.2mm]{sphere}} & \parbox[c]{1em} {\includegraphics[width=11.6mm]{dumbbell-155}} & 1.0 & 1.55 & 0.96 & 0.95\\
      & \textit{EV4} & \parbox[c]{1em}{\includegraphics[width=9.2mm]{sphere}} & \parbox[c]{1em} {\includegraphics[width=12.2mm]{dumbbell-17}} & 1.0 & 1.70 & 0.97 & 1.01\\
      \hline
      & \textit{VV1} & \parbox[c]{1em} {\includegraphics[width=10.2mm]{dumbbell-shape-size-11}} & \parbox[c]{1em} {\includegraphics[width=11.0mm]{dumbbell-shape-size-12}}  & 1.1 & 1.2 & 1.03 & 1.0\\
      \textit{Varying}& \textit{VV2}&  \parbox[c]{1em} {\includegraphics[width=10.2mm]{dumbbell-shape-size-11}} & \parbox[c]{1em} {\includegraphics[width=11.8mm]{dumbbell-shape-size-13}} & 1.1 & 1.3 & 1.06 & 1.03\\
       \textit{Volume} (\textit{VV}) & \textit{VV3}& \parbox[c]{1em} {\includegraphics[width=10.2mm]{dumbbell-shape-size-11}} &\parbox[c]{1em} {\includegraphics[width=12.5mm]{dumbbell-shape-size-14}} & 1.1 & 1.4 & 1.08 & 1.07\\
      & \textit{VV4}& \parbox[c]{1em}{\includegraphics[width=9.2mm]{sphere}} & \parbox[c]{1em} {\includegraphics[width=12.5mm]{dumbbell-shape-size-14}} & 1.0 & 1.4 & 1.12 & 1.04\\
      \hline
       & \textit{SS1}& \parbox[c]{1em}{\includegraphics[width=9.2mm]{sphere}} & \parbox[c]{1em} {\includegraphics[width=9.75mm]{sphere}} & 1.0 & 1.0 & 1.04 & 1.04\\
      \textit{Spheres}- & \textit{SS2} & \parbox[c]{1em}{\includegraphics[width=9.2mm]{sphere}} & \parbox[c]{1em} {\includegraphics[width=10.08mm]{sphere}} & 1.0 & 1.0 & 1.08 & 1.08\\
       \textit{Spheres} (\textit{SS}) & \textit{SS3}& \parbox[c]{1em}{\includegraphics[width=9.2mm]{sphere}} & \parbox[c]{1em} {\includegraphics[width=10.37mm]{sphere}} & 1.0 & 1.0 & 1.11 & 1.11\\
       & \textit{SS4}& \parbox[c]{1em}{\includegraphics[width=9.2mm]{sphere}} & \parbox[c]{1em} {\includegraphics[width=10.85mm]{sphere}} & 1.0 & 1.0 & 1.16 & 1.16\\
\end{tabular}
\end{ruledtabular}
\end{table}

Initially, homogeneous packings comprising equal number of species $A$ ($N=5880$) and $B$ ($N=5880$) (50\% volume of $B$) are prepared for \textit{EV} mixtures, as follows. A simple cubic lattice with alternating layers of species $A$ and $B$ is prepared. The particles are assigned random translational velocities in all three directions and then allowed to flow under gravity at $\beta=32^\circ$ for $t=100$ units, and then the inclination is reduced to 0$^\circ$. When the total kinetic energy of the system becomes of the order of 10$^{-5}$, we stop the simulation to obtain a homogeneous configuration. A similar procedure is followed for \textit{VV} and \textit{SS} mixtures, however, the mixtures contain a smaller number of species $B$ than $A$. Now, the particles are assigned random translational velocities in the range of -0.5 to 0.5 in all three directions keeping their positions unchanged. The inclination of the chute is set again at $\beta=32^\circ$ for an initial period to let the particles gain enough kinetic energy and is then 
reduced to the desired value ($\beta\in(21^\circ,31^\circ$)). The inclination angle for each mixture of each type is chosen such a way that the steady values of shear stress and pressure gradients are the same for all mixtures and the steady value of shear rate is high enough to yield a high effective friction. This is done using a trial and error process. Based on the study of \citet{guillard2016scaling}, this approach enables us to decouple the effect of particle shape and size on the net segregation force and the resulting extent of segregation.    

We monitor the total kinetic energy and the average vertical centroid position ($y_G$) of either of the species with time. When the profiles become steady, we divide the simulation control volume into small rectangular boxes (bins) of cross-sectional area 20$\times$20 and height 1 to calculate the average properties. The reported data are non-dimensionalized using appropriate combinations of $d$, $m$, and $g$ and are averaged over 4 sets with each set over a time duration of 250 units. 

For studying the dynamics of segregation, normally graded initial conditions are used. A simple cubic lattice with species $A$ occupying the lower half of the box and species $B$ in the upper half is prepared for \textit{EV} mixtures, and the particles are allocated random translational velocities as in the earlier case. A similar procedure is followed for \textit{VV} and \textit{SS} mixtures, but with species $B$ in the lower half. The mixture is then allowed to flow under gravity at the desired inclination angle. We note the average vertical centroid position ($y_G$) of species $A$ (or $B$) and the kinetic energy with time until they become constant. 

The mean velocity of species $A$ in the $x$-direction, $v^A_x$, in a bin is calculated as
\begin{equation}
 \boldsymbol{v}^A_x = \frac{1}{N^A}\sum\limits_{i=1}^{N^A}{\boldsymbol{c}^A_{xi}},
\end{equation}
where $c^A_{xi}$ is the instantaneous velocity of species $A$, and $N^A$ is the number of species $A$ in the bin. The volume fraction of species $A$, $\phi^A$, is computed as 
\begin{equation}
\phi^A = \frac{1}{N^A}\sum_{i=1}^{N^A} v^A_{pi}/V,
\end{equation}
where $v^A_{pi}$ is the volume of species $A$, and $V$ is the volume of the bin.
The mean velocity and volume fraction of species $B$ are calculated similarly. The total volume fraction, $\phi$, is computed as
\begin{equation}
\phi = \phi^A+\phi^B.
\end{equation}
The total stress tensor, composed of streaming and collisional components, is given as
\begin{equation}
\boldsymbol{\sigma}=\boldsymbol{\sigma}_s+\boldsymbol{\sigma}_c ,
\end{equation}
with the streaming stress given as \cite{Mandal2D}
\begin{equation}
\boldsymbol{\sigma}_s = \boldsymbol{\sigma}^A_s +\boldsymbol{\sigma}^B_s,
\end{equation}
with
\begin{equation}
\boldsymbol{\sigma_s}^A = \frac{m^A_p N^A}{V}\Big[\frac{1}{N^A}\sum\limits_{i=1}^{N^A}\boldsymbol{c}_i^A\boldsymbol{c}_i^A-\boldsymbol{v}^A\boldsymbol{v}^A\Big],
\end{equation} 
and the collisional stress as \cite{Mandal2D}
\begin{equation}\label{eq:anur}
\boldsymbol{\sigma}_c = \frac{1}{V} \sum_{c=1}^{N_c}{\boldsymbol{F}_{ij}\boldsymbol{x}_{ij}},
\end{equation}
where $\boldsymbol {F}_{ij}$ =  $\boldsymbol {F}_{nij}$ +  $\boldsymbol {F}_{tij}$, is the total force comprising normal and tangential components, exerted on particle $i$ by particle $j$, $\boldsymbol{x}_{ij}$ =  $\boldsymbol{x}_i$ - $\boldsymbol{x}_j$, is the position vector directing from the centroids of particles $j$ to $i$, and the summation is over all particle-particle contacts in the sampling volume ($V$).
The deviatoric stress tensor is given as 
\begin{equation}
\boldsymbol{\tau} = \boldsymbol{\sigma} + P\boldsymbol{I},
\end{equation}
where $\boldsymbol{I}$ is the unit tensor, and $P$ = -tr($\boldsymbol{\sigma}$)/3, is the pressure. The shear rate ($\dot{\gamma}=dv_x/dy$) is computed by differentiating the mean velocity profile using the forward differences.

\section{Results and discussion}\label{sec:results and discussion}
\subsection{Dynamics of segregation}
\begin{figure}
\centerline{\includegraphics[width=4.5in]{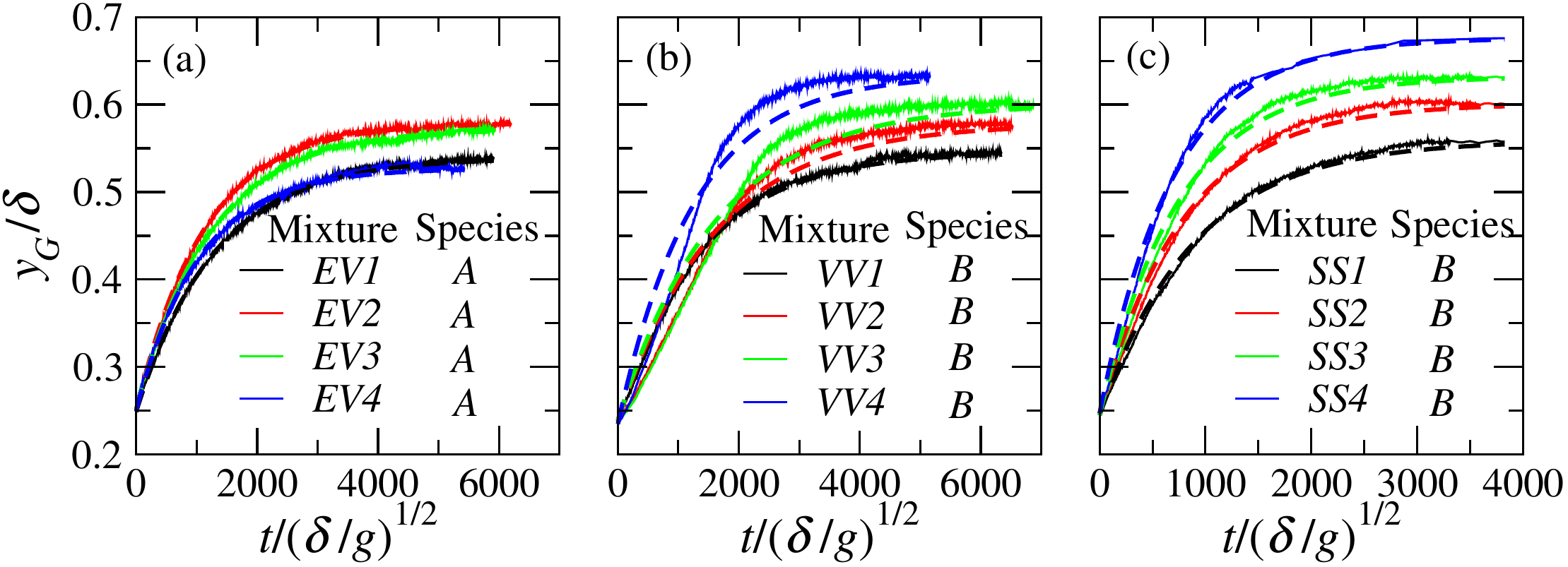}}
\caption{Variation of the normalized average vertical centroid position ($y_G/\delta$) of the species with normalized time ($t/\sqrt{\delta/g}$) for various mixtures: (a) Species $A$ for \textit{EV} mixtures, (b) species $B$ for \textit{VV} mixtures, and (c) species $B$ for \textit{SS} mixtures. The dashed lines are fits of Eq.~(\ref{eq:time}).}
\label{fig:timescale}
\end{figure}

\begin{figure}
\centerline{\includegraphics[width=3.5in]{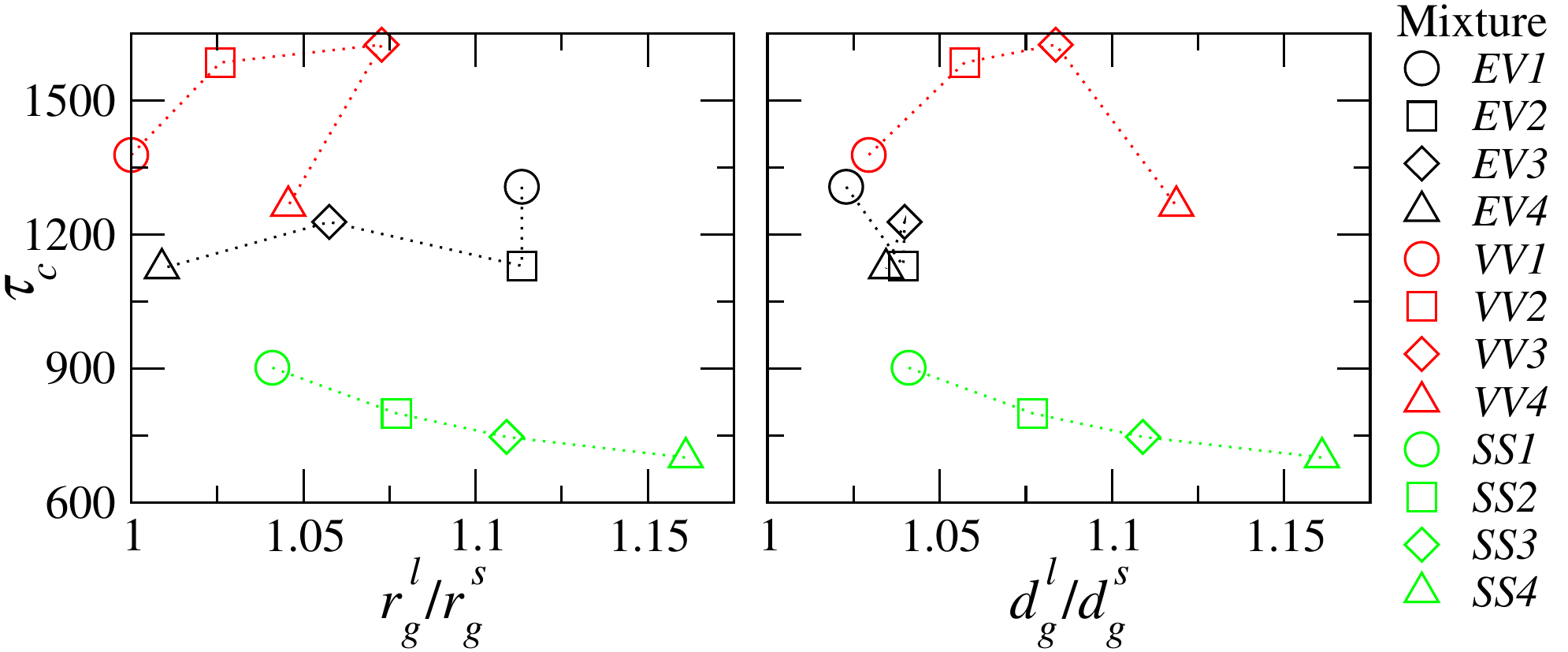}}
\caption{Variation of the characteristic time scale ($\tau_c$) with the effective size ratios, (a) $r^l_g/r^s_g$ and (b) $d^l_g/d^s_g$ with $l,s$ corresponding to the large and small species, respectively, for various mixtures, as indicated in the legend. The dotted lines are guide to the eye. }
\label{fig:tau}
\end{figure}

Fig.~\ref{fig:timescale} shows the evolution of segregation, in terms of the centroid position of one of the species, for the three types of mixtures: (1) \textit{EV}, (2) \textit{VV}, and (3) \textit{SS} (Table~\ref{tab:mixtures}). Species $A$ in \textit{EV} mixtures (Fig.~\ref{fig:timescale}(a)) and species $B$ in \textit{VV} mixtures (Fig.~\ref{fig:timescale}(b)) rise with time before reaching steady positions. Note that the species which rises up has a larger geometric mean diameter (effective size) in both cases. However, this is not always true if the radius of gyration is considered as the measure of effective size; species $A$ with smaller radius of gyration rises in mixture \textit{EV4}, and species $B$ shows an upward migration in mixture \textit{VV1}, though the ratio of the radii of gyration of the two species is one in this case. A steady value of $y_G/\delta$ $\approx0.75$, where $\delta$ is the flowing layer thickness, implies the complete segregation of the larger species in a pure layer near the free surface. Note that $\delta$ is taken to be the vertical distance from the base to the point, where the value of total volume fraction ($\phi$) is 50\% of bulk value. The segregation is incomplete ($y_G/\delta<0.75$) in all cases, as also seen from Fig.~\ref{fig:system}, due to the small geometric mean diameter ratios. The extent of segregation is highest in mixtures \textit{EV2} and \textit{VV4} and lowest in mixtures \textit{EV4} and \textit{VV1} for \textit{EV} and \textit{VV} mixtures, respectively. This is consistent with the size ratio. A similar dynamics of rise of larger spheres toward the free surface is noted for the \textit{SS} mixtures, and the larger sphere traverses more distance with increasing the size ratio, as expected (Fig.~\ref{fig:timescale} (c)). Following the works of \citet{dvziugys2009role}, \citet{hill2014segregation} and \citet{staron2014segregation}, we analyze the data by fitting the following equation       
\begin{equation}\label{eq:time}
y^\prime_G(t)= y^\prime_{G0} + (y^\prime_{Gf} - y^\prime_{G0})\Big[1-\exp(-\frac{t^\prime}{\tau_c})\Big],
\end{equation}
where $y^\prime_{G0}=y_{G0}/\delta$ and $y^\prime_{Gf}=y_{Gf}/\delta$ are the normalized initial and final values of $y_G$, respectively, and $t^\prime=t/\sqrt{\delta/g}$ is the normalized time. $\tau_c$ is the time scale pertaining to the rate of segregation. The fitted vales of $\tau_c$ are given in Fig.~\ref{fig:tau} as functions of the ratios of radii of gyration and geometric mean diameters. We note that the value of $\tau_c$ is the highest for \textit{VV} mixtures and the smallest for \textit{SS} mixtures for a given size ratio. $\tau_c$ monotonically decreases with increasing both size ratios for \textit{SS} mixtures, however, no correlation is obtained for \textit{EV} and \textit{VV} mixtures.

\subsection{Steady-state flow profiles}
\begin{figure}\vspace{0.0cm}
\centerline{\includegraphics[width=3.5in]{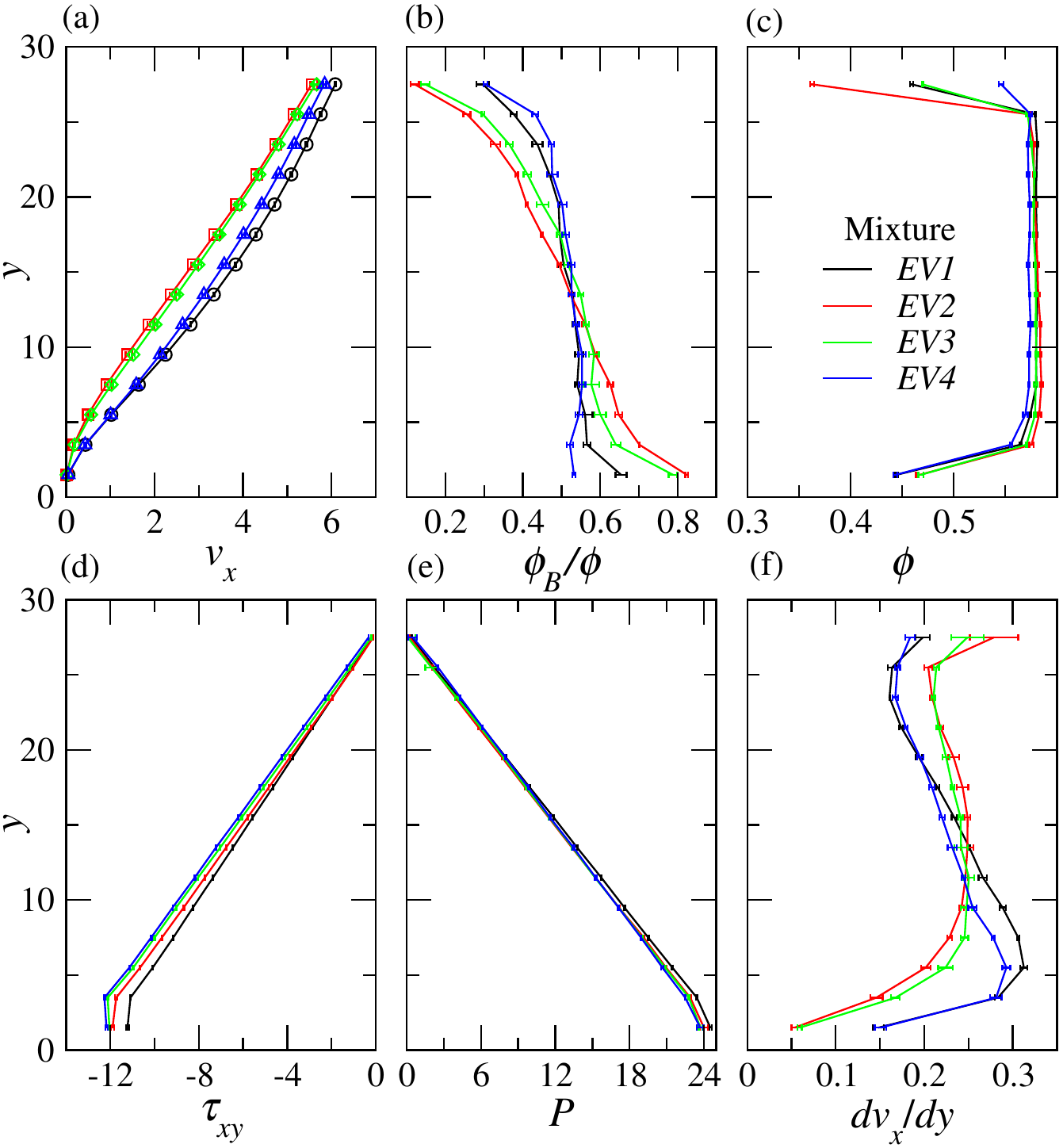}}
\caption{(a) Mean velocity ($v_x$) of species $A$ (symbols) and species $B$ (lines), (b) concentration of species $B$ ($\phi_B/\phi$), (c) total volume fraction ($\phi$), (d) shear stress ($\tau_{xy}$), (e) pressure ($P$), and (f) shear rate ($dv_x/dy$) profiles of the \textit{EV} mixtures, as indicated in the legend. The inclination angles for the four mixtures are $\beta=24^\circ$, 25.7$^\circ$, 26.4$^\circ$, and 26.7$^\circ$, respectively.}
\label{fig:shape-segregation}
\end{figure}
\begin{figure}[h!]\vspace{0.0cm}
\centerline{\includegraphics[width=3.5in]{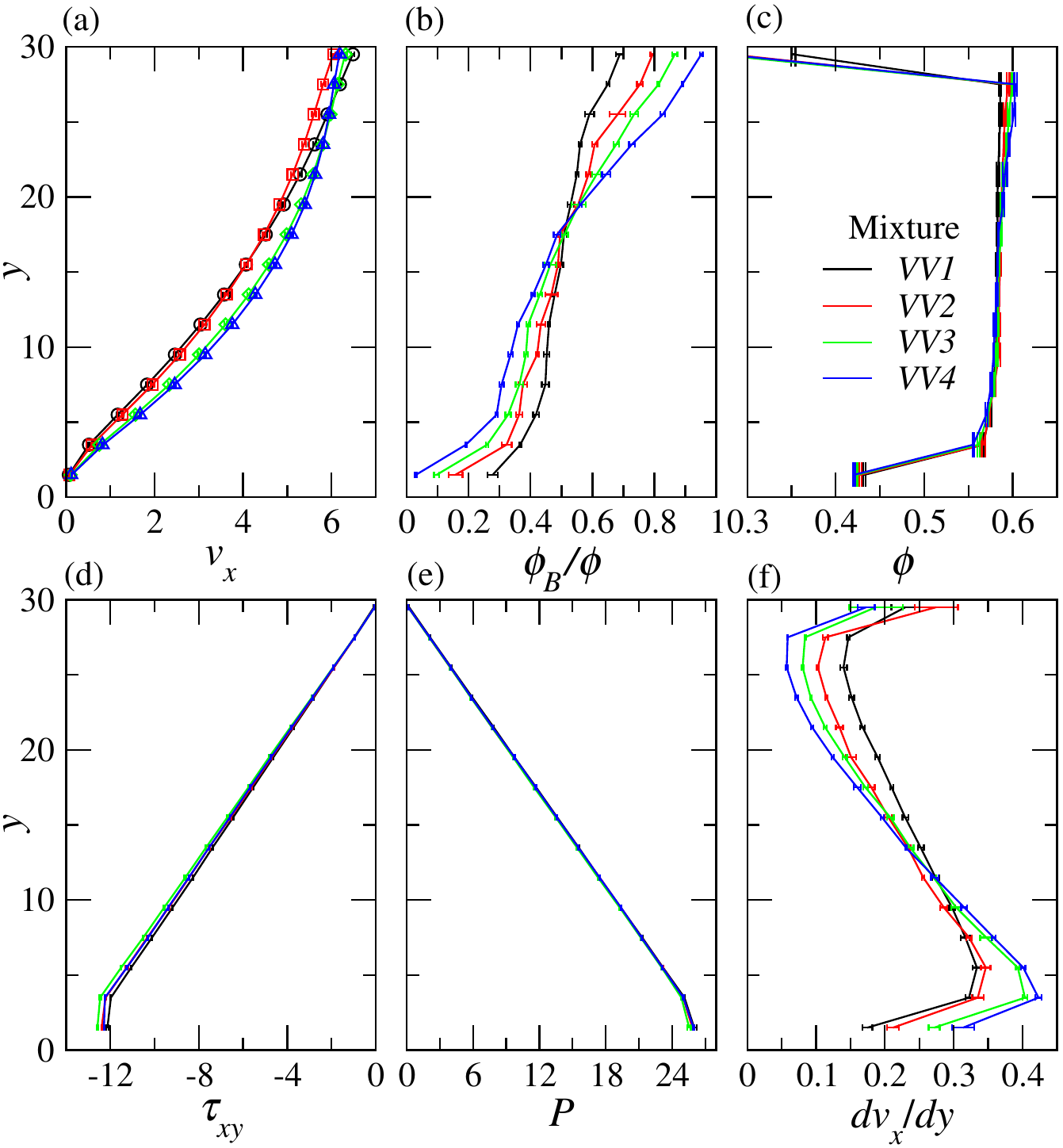}}
\caption{(a) Mean velocity ($v_x$) of species $A$ (symbols) and species $B$ (lines), (b) concentration of species $B$ ($\phi_B/\phi$), (c) total volume fraction ($\phi$), (d) shear stress ($\tau_{xy}$), (e) pressure ($P$), and (f) shear rate ($dv_x/dy$) profiles of the \textit{VV} mixtures, as indicated in the legend. The inclination angles for the four mixtures are $\beta=24.2^\circ$, 24.7$^\circ$, 25.2$^\circ$, and 24.6$^\circ$, respectively.}
\label{fig:shape-size-segregation}
\end{figure}
\begin{figure}\vspace{0.0cm}
\centerline{\includegraphics[width=3.5in]{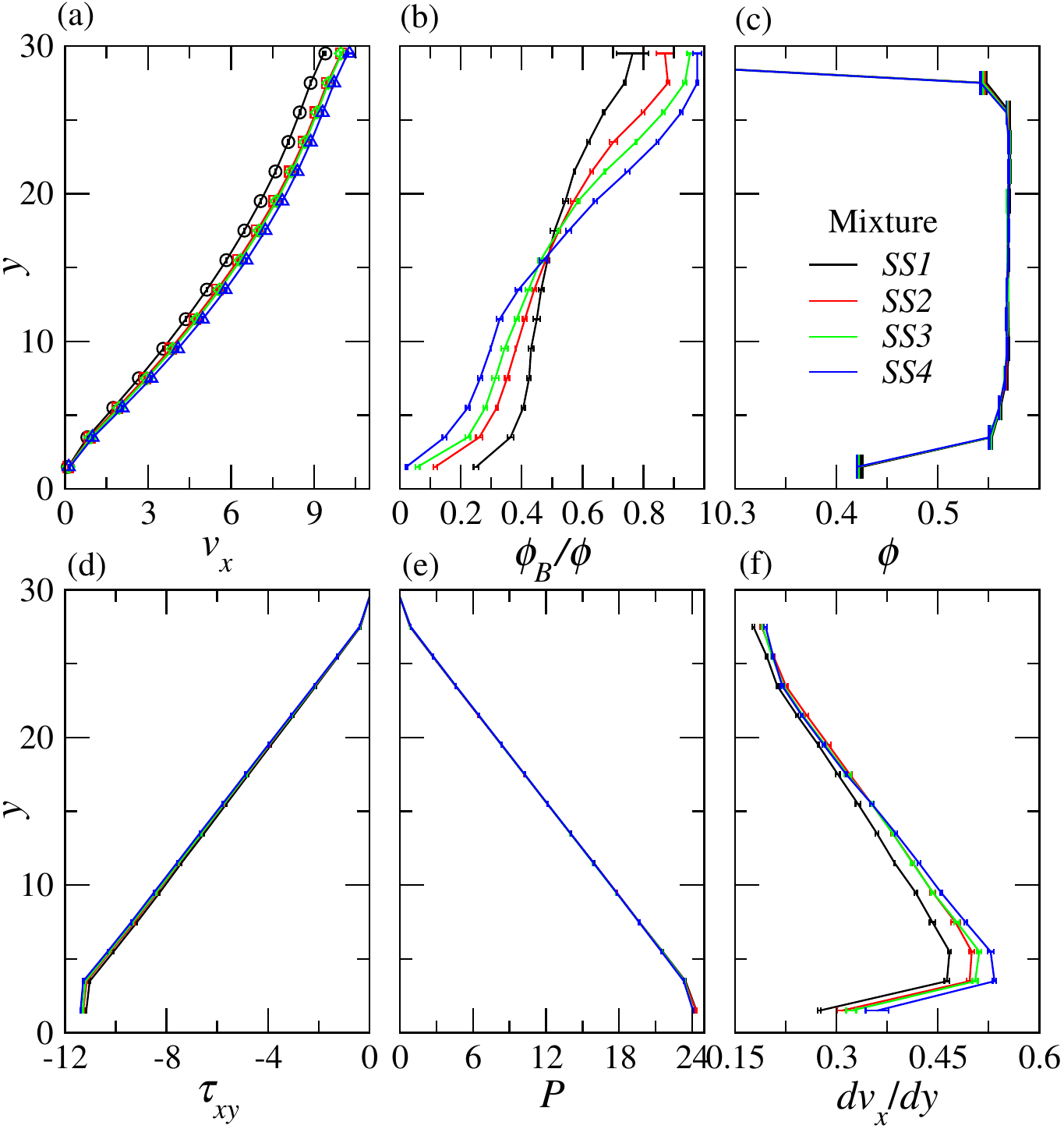}}
\caption{(a) Mean velocity ($v_x$) of species $A$ (symbols) and species $B$ (lines), (b) concentration of species $B$ ($\phi_B/\phi$), (c) total volume fraction ($\phi$), (d) shear stress ($\tau_{xy}$), (e) pressure ($P$), and (f) shear rate ($dv_x/dy$) profiles of the \textit{SS} mixtures, as indicated in the legend. The inclination angles for the four mixtures are $\beta=23.9^\circ$, 24.2$^\circ$, 24.3$^\circ$, and 24.5$^\circ$, respectively.}
\label{fig:size-segregation}
\end{figure}

Figs.~\ref{fig:shape-segregation}, \ref{fig:shape-size-segregation}, and \ref{fig:size-segregation} show the mean velocity ($v_x$), concentration of species $B$ ($\phi_B/\phi$), total volume fraction ($\phi$), shear stress ($\tau_{xy}$), pressure ($P$), and shear rate ($dv_x/dy$) profiles at steady state for \textit{EV}, \textit{VV}, and \textit{SS} mixtures, respectively. Although the maximum velocities for the four mixtures are nearly the same in each case, the profiles are different, and this is reflected in the shear rate profiles.
The mean velocities of species $A$ (symbols) and $B$ (lines) are the same for all mixtures in all cases. The concentration of species $B$ (dumbbells) is higher near the base ((Fig.~\ref{fig:shape-segregation}(b)) for all \textit{EV} mixtures, indicating again that species $B$ settles, and species $A$ floats up in these mixtures under shear flow. However, the concentration of species $B$ is higher near the free surface (Fig.~\ref{fig:shape-size-segregation}(b)) for all \textit{VV} mixtures, signifying a reverse scenario of the migration of the species in comparison with the earlier case. The steady-state concentration profiles are unaffected by the initial condition; the mixtures exhibit the same concentration profiles with the normally graded initial conditions, as with the homogeneous initial conditions used here. The geometric mean diameter of species $B$ is smaller in \textit{EV} mixtures and larger in \textit{VV} mixtures than that of species $A$, which validates the direction of segregation, but this not always true for the case of radius of gyration. Hence, the direction of segregation in a shear flow is not consistent with the results for vibrated systems \cite{roskilly2010investigating}. The spheres with larger sizes accumulate near the free surface (Fig.~\ref{fig:size-segregation}(b)) in \textit{SS} mixtures, as expected.     

The degree of segregation may also be quantified by the slope $d(\phi_B/\phi)/dy$ (Figs.~\ref{fig:shape-segregation}(b),\ref{fig:shape-size-segregation}(b),\ref{fig:size-segregation}(b)) at the middle of the flowing layer: $d(\phi_B/\phi)/dy=0$ implies the perfect mixing, and $d(\phi_B/\phi)/dy=\infty$ implies the perfect segregation. The degree of segregation monotonically increases with increasing the geometric mean diameter ratio for \textit{VV} and \textit{SS} mixtures, though it decreases and then increases for \textit{EV} mixtures. The degrees of segregation for all \textit{EV} mixtures are smaller than those for \textit{VV} and \textit{SS} mixtures, which is again consistent with the geometric mean diameter ratio. With increasing angle, the mean velocity, shear rate, and shear stress increase, the pressure remains almost constant, and the total volume fraction decreases (not shown) for all the mixtures in all cases. The degree of segregation (given by $d(\phi_B/\phi)/dy$) decreases with increasing angle for \textit{VV} and \textit{SS} mixtures, though it remains almost constant for \textit{EV} mixtures. With increasing concentration of species $B$ in the mixtures, the local concentration of species $B$ increases near the base for \textit{EV} mixtures and the free surface for \textit{VV} and \textit{SS} mixtures, and the thickness of a mixed layer increases in the middle (not shown). 
 
 \begin{figure}\vspace{0.0cm}
\centerline{\includegraphics[width=4.5in]{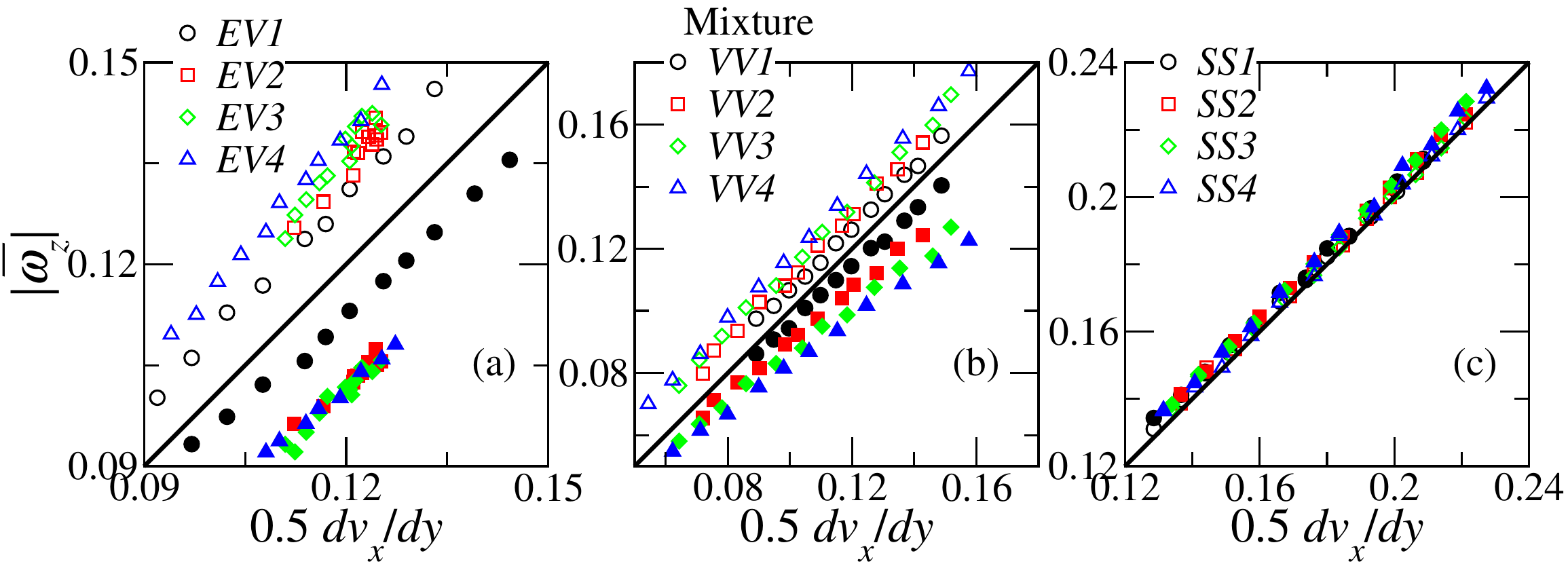}}
\caption{The absolute mean rotational velocity ($|\overline{\omega}_z|$) of species $A$ (open symbols) and species $B$ (filled symbols) in the vorticity direction as a function of the shear rate ($dv_x$/$dy$) for various types of mixtures (a) \textit{EV}, (b) \textit{VV}, and (c) \textit{SS}, respectively.}
\label{fig:rot}
\end{figure}
Figure~\ref{fig:rot} shows the variation of the angular velocity of the different species versus the local continuum angular velocity obtained from the vorticity. The rotational velocity of species $B$ (filled symbols) is smaller than that of species $A$ (open symbols) for all \textit{EV} and \textit{VV} mixtures (Figs.~\ref{fig:rot}(a),(b)) and the local continuum angular velocity, denoted by the diagonal lies between them. This is a consequence of the alignment of the dumbbells in the flow direction as seen previously \cite{Mandal2D,Mandal3D}. In the case of mixtures of spheres (\textit{SS}), the rotational velocities of the two species are equal and nearly equal to the local continuum angular velocity (Fig.~\ref{fig:rot}(c)), as expected.

We compare the steady-state concentration profiles for various mixtures with the predictions of a continuum theory \cite{gray2006particle} developed for gravity-driven free surface flows of binary mixtures of spheres next. The theoretical non-dimensionalized concentration profile of a species is given by \cite{gray2006particle}
\begin{equation}\label{eq:f}
f(y^\star) = \frac{\big(1-\exp(-\bar{f}Pe)\big)\exp\big(\pm(\bar{f}-y^\star)Pe\big)}{1-\exp\big(-(1-\bar{f})Pe\big)+\big(1-\exp(-\bar{f}Pe)\big)\exp\big(\pm(\bar{f}-y^\star)Pe\big)},
\end{equation}
where $f=\phi^{B}/\phi$ is the concentration of species $B$, $y^\star=y/\delta$, $\bar{f}=\int_0^1 fdy^\star$ (0.5 in the present case) is the mean concentration over the flowing layer, and $Pe=q/D_r$ is the P$\acute{e}$clet number with $q$ and $D_r$ being the non-dimensionalized segregation velocity and diffusivity, respectively. $Pe=0$ quantifies no segregation and $Pe=\infty$, complete segregation with pure layers of large and small species near the free surface and the base \cite{gray2006particle}. In Eq.~\ref{eq:f}, ``--" is taken for large species, and ``+" for small species.

\begin{figure}\vspace{0.0cm}
\centerline{\includegraphics[width=4.5in]{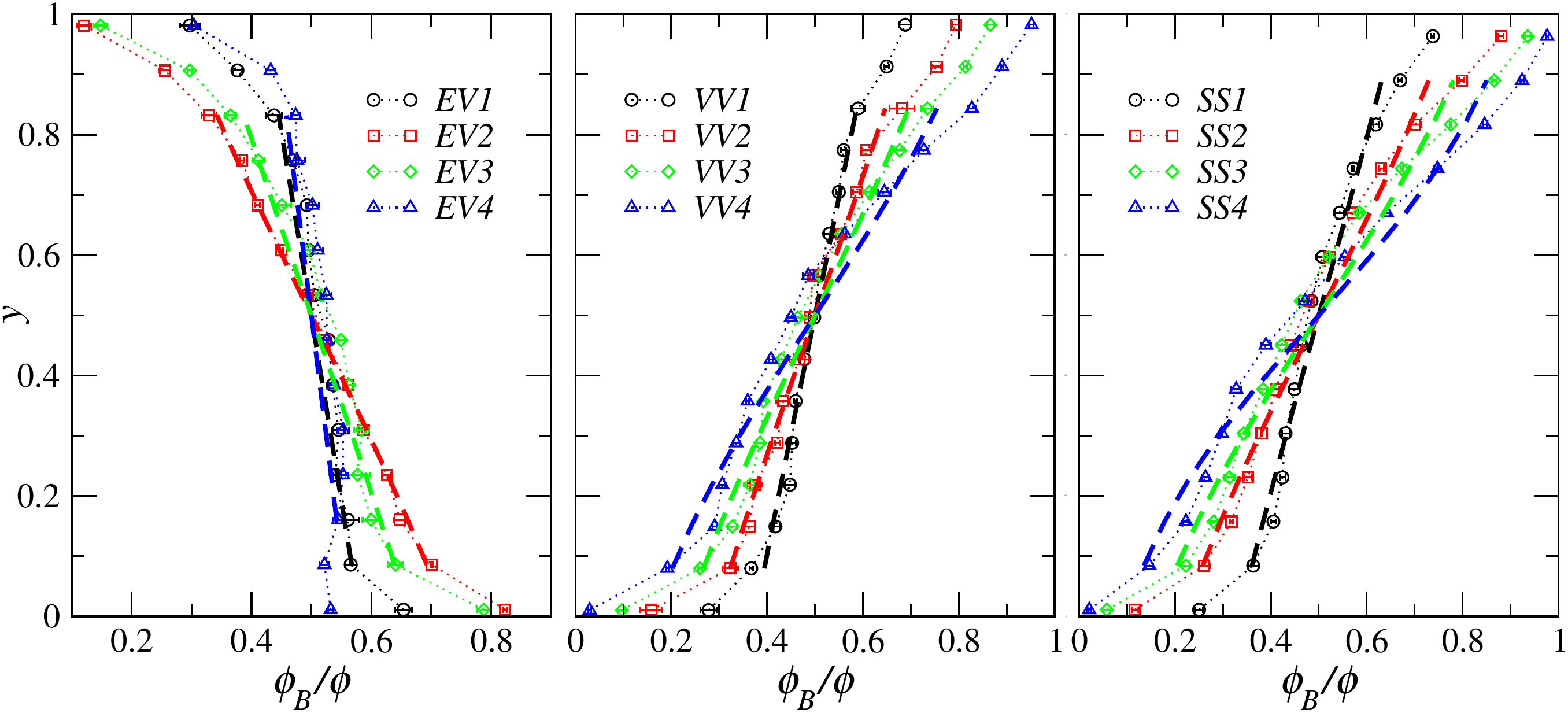}}
\caption{Comparisons of the steady-state concentration profiles (symbols) with the theoretical predictions of Eq.~(\ref{eq:f}) (dashed lines) for various types of mixtures (a) \textit{EV}, (b) \textit{VV}, and (c) \textit{SS}, respectively.}
\label{fig:theo}
\end{figure}

Fig.~\ref{fig:theo} shows a comparison of the predictions of Eq.~(\ref{eq:f}) with the concentration profiles obtained from the simulations for different mixtures. $Pe$ is taken as a fitting parameter. \citet{gray2006particle} showed that the theory is valid only in the dense flowing region. Hence, we consider the data only in the dense flowing region. The concentration profiles are in good agreement with the theoretical predictions (dashed lines) for all mixtures, implying the validity of the continuum theory for the mixtures of spherical and nonspherical particles. The values of $Pe$ are given in the next section. 

\subsection{Measures of segregation}
\begin{figure}[hb!]
\centerline{\includegraphics[width=4.5in]{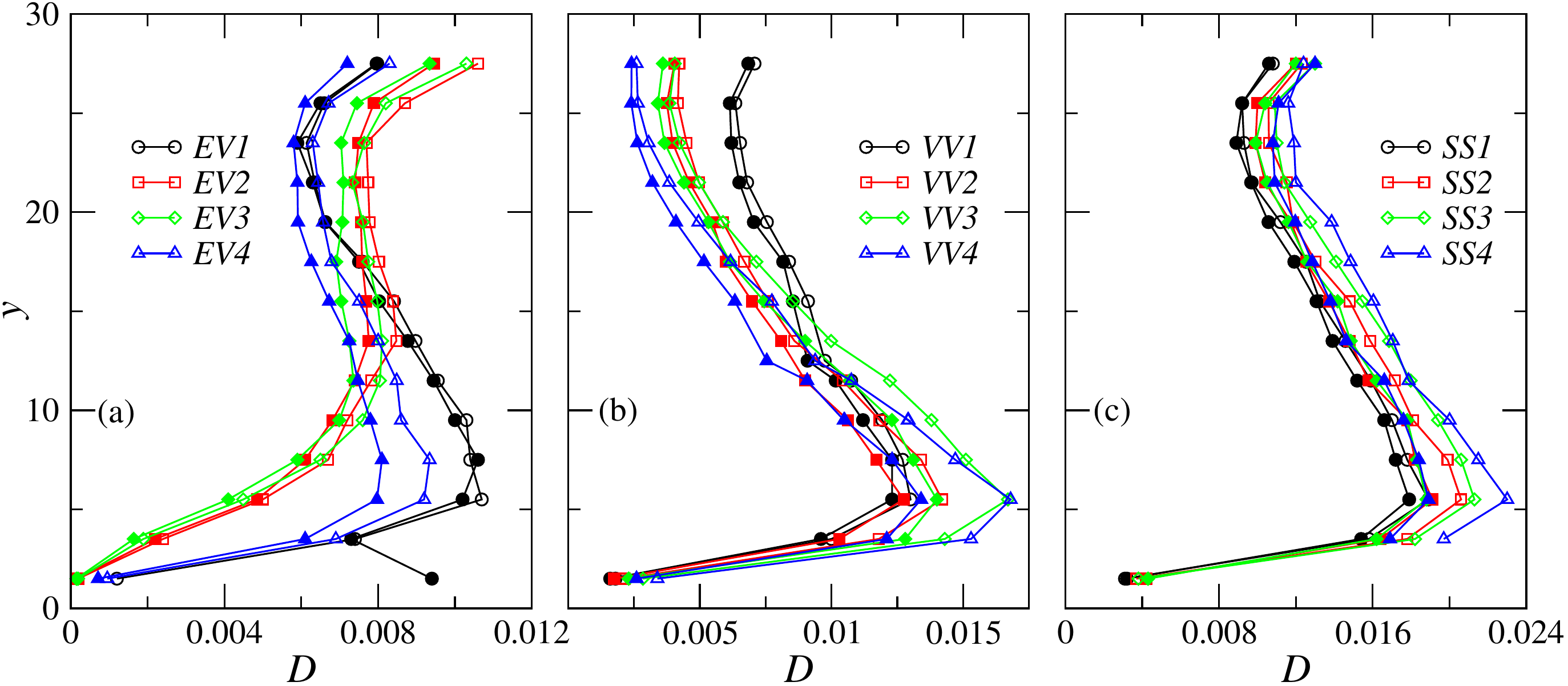}}
\caption{The variation of the self-diffusivity ($D$) of species $A$ (open symbols) and species $B$ (filled symbols) with the flowing layer height ($y$) for various types of mixtures (a) \textit{EV}, (b) \textit{VV}, and (c) \textit{SS}, respectively.}
\label{fig:D}
\end{figure}

We also measure the extent of segregation in terms of the segregation index ($\xi_s$) and segregation flux ($S$). We define $\xi_s$ as
\begin{equation}
\xi_s = \frac{(y_{Gf}-y_{G0})}{\delta}. 
\end{equation}
$\xi_s\approx0$ corresponds to no segregation (perfect mixing), and $\xi_s\approx0.5$ corresponds to complete segregation. 
 The segregation flux is equal to the diffusive flux at equilibrium \cite{ottino2000mixing,gray2006particle,fan2014modelling}. We estimate the segregation flux as 
\begin{equation}\label{eq:segre}
S = |D \frac{d(\phi_B/\phi)}{dy}|, 
\end{equation}  
where $D=\big(\phi^A D^A + \phi^B D^B\big)/(\phi^A + \phi^B)$ is the local volume averaged diffusivity, and $D^A$ and $D^B$ are the self-diffusivities of species $A$ and $B$ in the $y$-direction. Note that $D$ is depth-averaged over the dense flowing region to compute $S$ from Eq.~(\ref{eq:segre}). $D^A$ and $D^B$ are computed from the simulations as half the slope of the fitted lines to the mean square displacement \textit{versus} time plots \cite{Mandal3D}, and the values are given in Fig.~\ref{fig:D} for different mixtures. The diffusivity profile is similar to the shear rate profile (shown above) for each mixture in agreement with the scaling proposed by \citet{savage1993studies}. The profiles of the two species are similar to each other, however, their magnitudes are different. 

\begin{figure}[ht!]
\centerline{\includegraphics[width=6in]{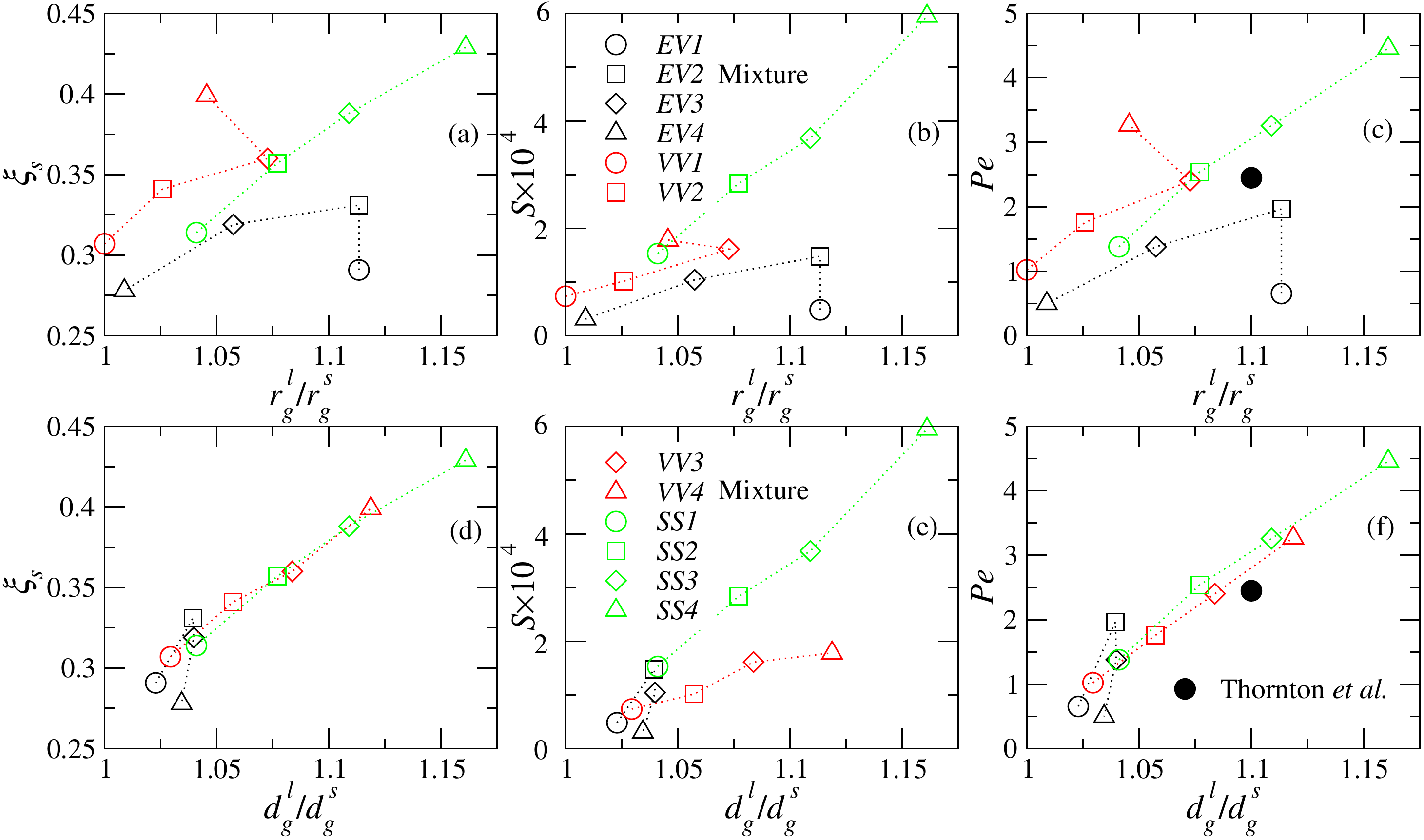}}
\caption{Variation of the (a),(d) segregation index ($\xi_s$), (b),(e) segregation flux ($S$), and (c),(f) P$\acute{e}$clet number ($Pe$) with the effective size ratios ($r^l_g/r^s_g$ or $d^l_g/d^s_g$ with $l,s$ corresponding to the large and small species, respectively) for various mixtures, as indicated in the legend. The dotted lines are guide to the eye. The filled symbol corresponds to the data taken from the study of \citet{thornton2012modeling}.}
\label{fig:flux}
\end{figure}

A comparison of the segregation index, segregation flux, and P$\acute{e}$clet number as functions of the ratios of radii of gyration $r^l_g$/$r^s_g$ and geometric mean diameters $d^l_{g}$/$d^l_{g}$ with $l,s$ corresponding to the large and small species, respectively, is shown in Fig.~\ref{fig:flux} for all the mixtures. 
The variation of $\xi_s$, $S$, and $Pe$ is poorly correlated to the variation of the ratio of radii of gyration (Fig.~\ref{fig:flux}(a),(b),(c)); the values of $\xi_s$, $S$ and $Pe$ are non-zero and finite for mixture \textit{VV1}, though the ratio $r^l_g/r^s_g$ is equal to one. A significant scatter of the data is also
present in $\xi_s$ \textit{versus} $r^l_g/r^s_g$, $S$ \textit{versus} $r^l_g/r^s_g$, and $Pe$ \textit{versus} $r^l_g/r^s_g$ plots. 
$\xi_s$, $S$, and $Pe$ increase monotonically with increasing geometric mean diameter ratio for all types of mixtures. The segregation index and P$\acute{e}$clet number correlate reasonably well with the ratio of geometric mean diameters (Fig.~\ref{fig:flux}(d),(f)), showing a good collapse of the data for all the mixtures. However, the segregation fluxes for the \textit{VV} mixtures are smaller in comparison with the \textit{EV} and \textit{SS} mixtures (Fig.~\ref{fig:flux}(e)). P$\acute{e}$clet number, obtained in the study \cite{thornton2012modeling} of segregation of an equal volume binary granular mixture of spheres with a size ratio of 1.1 on a rough chute inclined at 25$^\circ$, closely matches with our results. \citet{jing2017micromechanical} recently showed that if the rotation of the larger spheres in an equal volume binary granular mixture of spheres was suppressed by reducing the inter-particle friction, the extent of segregation decreased significantly. However, the reduction of the extent of segregation due to the smaller tendency of rotation of species $B$ does not quite happen in the \textit{EV} and \textit{VV} mixtures, which is due to the fact that the difference between the rotational velocity of species $B$ and the continuum angular velocity is quite small.    
\subsection{Rheology}
Two empirical constitutive laws: friction and dilatancy are used to describe the rheology of monodisperse spherical/nonspherical particles \cite{jop2006constitutive,Mandal2D,Mandal3D}. We follow the approaches of \citet{rognon2007dense} and \citet{tripathi2011rheology} to extend the constitutive laws for the mixtures. The friction law is given as 
\begin{equation}\label{eq:friction}
\mu(I)=\frac{|\tau_{xy}|}{P}= \mu_s + \frac{(\mu_m - \mu_s)}{(1 + I_0/I)},
\end{equation}
where $\mu_{s,m}=\big(\phi^A \mu_{s,m}^A + \phi^B \mu_{s,m}^B\big)/(\phi^A + \phi^B)$ and $I_0=\big(\phi^A I_0^A + \phi^B I_0^B\big)/(\phi^A + \phi^B)$ are local volume averaged model parameters with $\mu_{s,m}^{A,B}$ and $I_0^{A,B}$ be the model parameters of pure species $A,B$. $I = \dot{\gamma} d^{mix}/(P/\rho_p)^{1/2}$ is the generalized inertial number with $d^{mix} = (\phi^A d^A + \phi^B d^B)/ (\phi^A + \phi^B)$. The dilatancy law is given as 
\begin{equation}\label{eq:dilation}
\phi(I) = \phi_{max} - b{I}^n,
\end{equation}
where $\phi_{max}=\big(\phi^A \phi_{max}^A + \phi^B \phi_{max}^B\big)/(\phi^A + \phi^B)$, $b=\big(\phi^A b^A + \phi^B b^B\big)/(\phi^A + \phi^B)$, and $n=\big(\phi^A n^A + \phi^B n^B\big)/(\phi^A + \phi^B)$
are local volume averaged model parameters with $\phi_{max}^{A,B}$, $b^{A,B}$, $n^{A,B}$ be the model parameters of pure species $A,B$. Note that the above formulation for the mixture simplifies to that for the monodisperse particles when one of the species is absent, i.e., $\phi^A$ or $\phi^B$ is zero.
\begin{figure}[ht!]
\centerline{\includegraphics[width=5.9in]{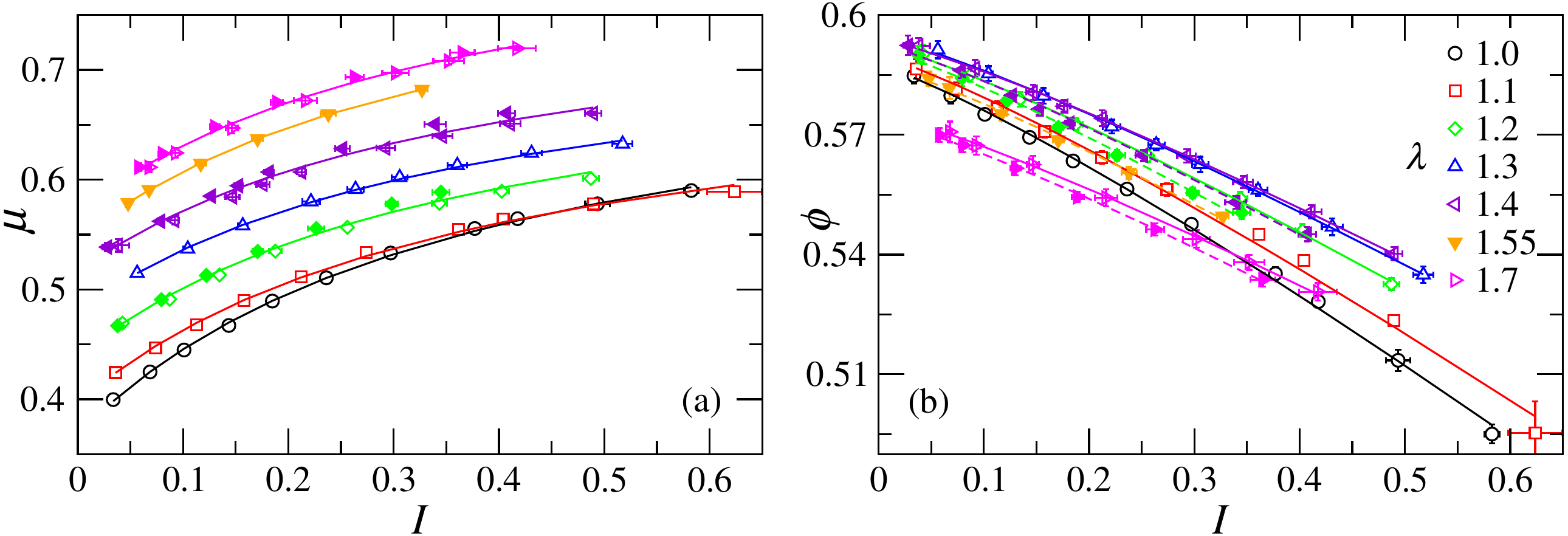}}
\caption{Variation of (a) the effective friction coefficient ($\mu$) and (b) the volume fraction ($\phi$) with inertial number ($I$) for monodisperse particles of different aspect ratios. Open symbols: data for species with fixed component diameter; filled symbols: data for species with varying component diameter in Table~\ref{tab:dumbbells}. The lines are fits of Eqs.~(\ref{eq:friction}) in (a) and (\ref{eq:dilation}) in (b).}
\label{fig:mu-I}
\end{figure} 
As mentioned before, Eqs.~(\ref{eq:friction}), (\ref{eq:dilation}) are well-posed only in the dense flow regime for moderate values of inertial number \cite{barker2015well}.

\begin{figure}[ht!]
\centerline{\includegraphics[width=4.5in]{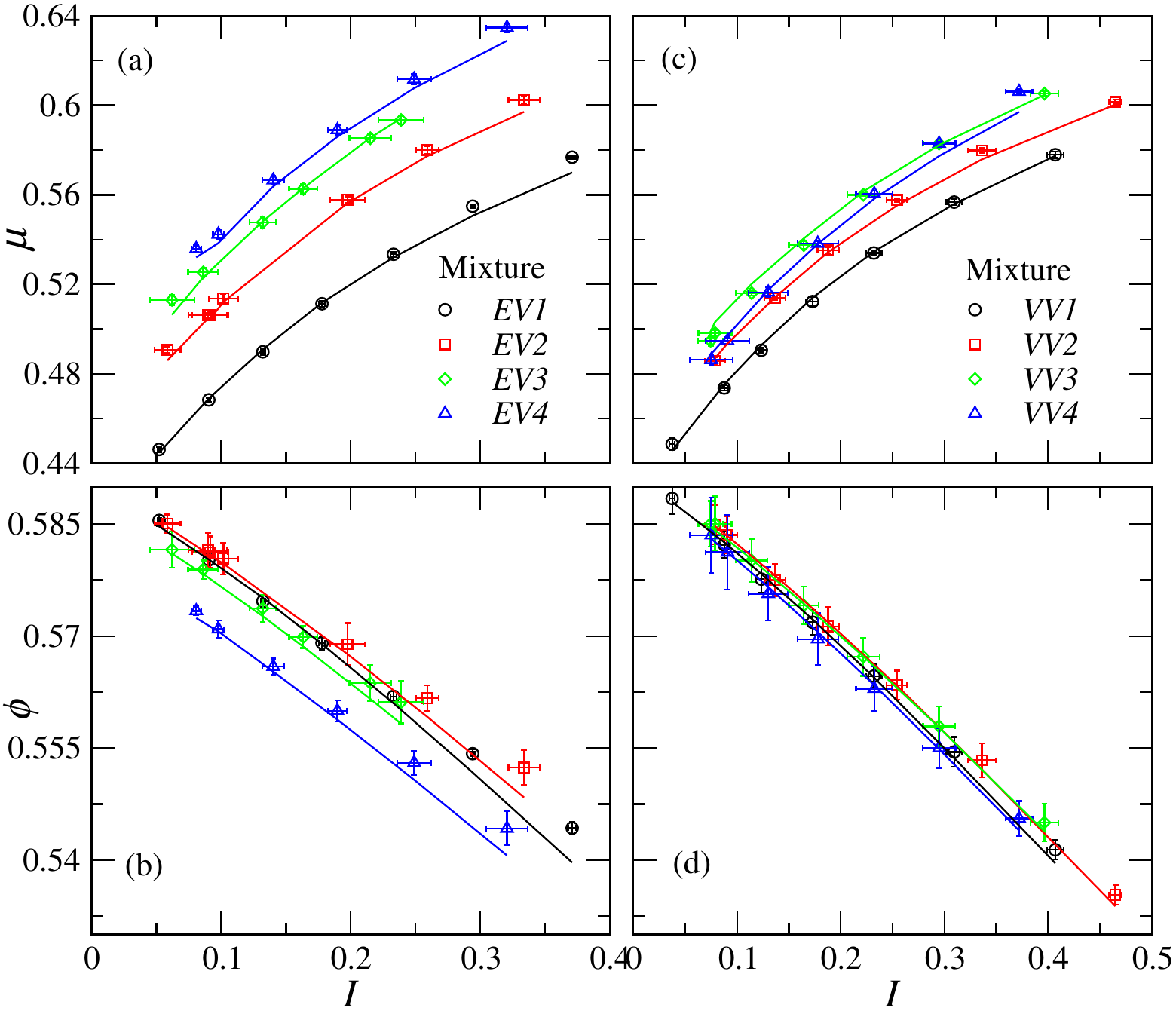}}
\caption{Variation of (a),(c) the effective friction coefficient ($\mu$) and (b),(d) the volume fraction ($\phi$) with generalized inertial number ($I$) for various mixtures, as indicated in the legend. The lines are predictions of Eq.~(\ref{eq:friction}) in (a) and (c) and Eq.~(\ref{eq:dilation}) in (b) and (d).}
\label{fig:mu-I-mix}
\end{figure}

We run simulations for monodisperse particles (particles of pure $A$ or $B$) of different aspect ratios at different angles and plot the depth-averaged $\mu$ \textit{versus} $I$ and $\phi$ \textit{versus} $I$ data (symbols) over the dense flowing region at each angle in Fig.~\ref{fig:mu-I}. Eqs.~(\ref{eq:friction}) and (\ref{eq:dilation}) are fitted to obtain the model parameters. The effective friction coefficient ($\mu$) and the volume fraction ($\phi$) increase significantly with increasing aspect ratio ($\lambda$) for a given inertial number ($I$) \cite{Mandal2D,Mandal3D}. The theoretical fits (lines) are quite good for both $\mu$ \textit{versus} $I$ and $\phi$ \textit{versus} $I$ plots with the fitted model parameters given in Table~\ref{tab:parameters}. The model parameters of the friction law are found to be the same for the dumbbells with different component diameters for a given aspect ratio; however, the model parameters of the dilatancy law are slightly different. We similarly compute the depth-averaged $\mu$ \textit{versus} $I$ and $\phi$ \textit{versus} $I$ data over the dense flowing region at each angle and plot them in Fig.~\ref{fig:mu-I-mix} for various mixtures. We find a good agreement between the simulation data (symbols) and theoretical prediction (lines) for all \textit{EV} and \textit{SS} mixtures, showing the validity of the constitutive relations for the mixtures of spherical and nonspherical particles. A similar agreement between the theory and simulations is found for \textit{SS} mixtures (not shown).
\begin{table}
\caption{\label{tab:parameters} The fitted model parameters.}
\begin{ruledtabular}
\begin{tabular}{lccccccc}
 Case  & $\lambda$ & $\mu_s$ & $\mu_m$ & $I_0$ & $\phi_{max}$ & $b$ & $n$ \\[1pt]
   \hline
   & 1.0 & 0.37 & 0.74 & 0.38 & 0.59 & 0.17 & 1.18\\
varying  & 1.2 & 0.44 & 0.74 & 0.38 & 0.59 & 0.15 & 1.18\\
 component   & 1.4 & 0.52 & 0.78 & 0.38 & 0.59 & 0.14 &1.18\\
 diameter & 1.55 & 0.55 & 0.84 & 0.38 & 0.59 & 0.14 &1.18\\
  & 1.7 & 0.57 & 0.86 & 0.38 & 0.57 & 0.13 & 1.18\\
  \hline
   & 1.0 & 0.37 & 0.74 & 0.38 & 0.59 & 0.17 & 1.18\\
  fixed & 1.1 & 0.40 & 0.72 & 0.38 & 0.59 & 0.16 & 1.18 \\
  component & 1.2 & 0.44 & 0.74 & 0.38 & 0.59 & 0.14 & 1.18 \\
  diameter & 1.3 & 0.48 & 0.75 & 0.38 & 0.595 & 0.13 & 1.18\\
  & 1.4 & 0.52 & 0.78 & 0.38 & 0.59 & 0.125 & 1.18\\ 
\end{tabular}
\end{ruledtabular}
\end{table}
\section{Conclusions}\label{sec:conclusions}
We studied the flow of granular mixtures down a rough inclined plane using three types of mixtures --- (1) \textit{EV} mixtures of species with the same volume but different aspect ratios, (2) \textit{VV} mixtures of species with different volumes and different aspect ratios, and (3) \textit{SS} mixtures of spheres with the same volume ratios as in (2). The inclination angle of the base for each mixture was adjusted and maintained at a high value to yield the same pressure and shear stress gradients for all mixtures and a high effective friction for each. Following \citet{guillard2016scaling}, this ensured that the segregation force and resulting extent of segregation were a function of only the size and shape of the particles. The radius of gyration and geometric mean diameter were used as two size parameters for the particles. The species with larger geometric mean diameter migrated to the free surface in all three types of mixtures. However, this rule did not always apply for the radius of gyration, contrary to the results obtained for vibrated systems \cite{roskilly2010investigating}. This indicates that the mechanisms of shape induced segregation in shear flows are different from those in vibrated systems. The rate of segregation was highest for the \textit{SS} mixtures for a given size ratio. At steady state, the concentration of species with larger geometric mean diameter was higher near the free surface in all three types of mixtures. The concentration profiles obtained from the simulations were in good agreement with the continuum theory of \citet{gray2006particle}. The segregation index ($\xi_s$), segregation flux ($S$), and P$\acute{e}$clet number ($Pe$) were poorly correlated with the ratio of radii of gyration. However, $\xi_s$ and $Pe$ were well correlated with the ratio of geometric mean diameters, showing a near-perfect collapse of the data for all the mixtures considered and increased monotonically with increasing geometric mean diameter ratio for all types of mixtures. The result of \citet{thornton2012modeling} for the mixtures of different size spheres also agreed with the results of the $Pe$ versus ratio of geometric mean diameters from our simulations. The results indicate that the geometric mean diameter is a good measure of the effective size for nonspherical particles. The approach followed in the study will be useful in quantifying segregation of mixtures comprising complex shaped particles. The extended $\mu-I$ and $\phi-I$ scaling relations were shown to be applicable for describing the rheology of the mixtures of spherical and nonspherical particles with the model parameters obtained as volume averages of the pure component values of the different species.

\begin{acknowledgments}
The authors acknowledge the financial support of the Science and Engineering Research Board, India through grant SR/S2/JCB-34/2010.
\end{acknowledgments}

\providecommand{\noopsort}[1]{}\providecommand{\singleletter}[1]{#1}%
\end{document}